\documentclass[a4paper,fleqn,usenatbib,useAMS]{mnras}

\pdfoutput=1
\usepackage[T1]{fontenc}
\usepackage{aecompl}

\usepackage[pdftex]{graphicx}	
\usepackage{amsmath}	
\usepackage{amssymb}	
\usepackage{threeparttable}
\usepackage{bm}		
\usepackage{pdflscape}	

\newcommand{\solmass}{\rm M_{\odot}}
\newcommand{\solmetal}{\rm Z_{\odot}}

\providecommand{\adsurl}[1]{\href{#1}{ADS}}
 
\newcommand{\electron}{\rm e^-}
\newcommand{\Hp}{\rm H^+}
\newcommand{\Hm}{\rm H^-}
\newcommand{\HH}{\rm H_2}
\newcommand{\HHp}{\rm H_2^+}
\newcommand{\per}[2]{\rm {#1}^{-#2}}
\newcommand{\diff}[3]{\mathinner{\frac{d ^{#1} {#2}}{d {#3} ^{#1}}}}

\title[Formation of globular clusters II]
{Formation of globular clusters induced by external ultraviolet radiation II: Three-dimensional radiation hydrodynamics simulations}
\author[M. Abe, M. Umemura, and K. Hasegawa]{Makito Abe$^{1}$\thanks{E-mail: mabe@ccs.tsukuba.ac.jp (MA)}, Masayuki Umemura$^{1}$
and Kenji Hasegawa$^{2}$ \\
$^{1}$Center for Computational Sciences, University of Tsukuba, Ten-nodai, 1-1-1 Tsukuba, Ibaraki 305-8577, Japan\\
$^{2}$Graduate School of Science, Nagoya University, Furo-cho, Chikusa-ku, Nagoya, Aichi 464-8602, Japan}

\date{Accepted .Received ; in original form}

\pubyear{2016}

\begin{document}
\label{firstpage}
\pagerange{\pageref{firstpage}--\pageref{lastpage}}
\maketitle

\begin{abstract}
We explore the possibility of the formation of globular clusters under ultraviolet (UV) background radiation.
One-dimensional spherical symmetric radiation hydrodynamics (RHD) simulations by Hasegawa et al. have demonstrated 
that the collapse of low-mass ($10^{6-7}\solmass$) gas clouds exposed to intense UV radiation 
can lead to the formation of compact star clusters like globular clusters (GCs) 
if gas clouds contract with supersonic infall velocities. 
However, three-dimensional effects, such as the anisotropy of background radiation 
and the inhomogeneity in gas clouds, have not been studied so far. 
In this paper, we perform three-dimensional RHD simulations in a semi-cosmological context, and
reconsider the formation of compact star clusters in strong UV radiation fields.
As a result, we find that although anisotropic radiation fields bring an elongated shadow
of neutral gas, almost spherical compact star clusters can be procreated from a ``supersonic infall'' cloud,
since photo-dissociating radiation suppresses the formation of hydrogen molecules in the shadowed regions
and the regions are compressed by UV heated ambient gas. 
The properties of resultant star clusters match those of GCs. 
On the other hand, in weak UV radiation fields, dark matter-dominated star clusters 
with low stellar density form due to the self-shielding effect as well as the positive feedback by ionizing photons. 
Thus, we conclude that 
the ``supersonic infall'' under a strong UV background is a potential mechanism to form GCs.
\end{abstract}



\section{Introduction}
According to the concordant cosmology, the formation of low-mass sub-galactic objects are 
thought to have been the prime mode of the star formation in the early Universe. 
Considering the fact that stars are born in the form of star clusters in present-day galaxies 
\citep[e.g.,][]{Lada&Lada2003,Meurer+95,Fall+05},
it is of great importance to explore the formation of star clusters in such sub-galactic objects, to reveal the structure formation history in the Universe. 
Globular clusters (GCs) are significant tracers of early star formation history, since
they are low-metal, oldest star clusters in the Universe. 
GCs are relatively massive ($10^{4-6}\solmass$) and stellar-dominated systems 
in which stars are tightly distributed in color-magnitude diagram. 
Thus, GCs are thought to be of a single stellar population. Their ages can be evaluated by isochrone fitting. 
Although there are some uncertainties in the distances to GCs, the metallicity, and the stellar evolution models, 
the typical age is evaluated to be $\gtrsim 10$ Gyr with an uncertainty of $\sim$ Gyr \citep[e.g.,][]{Krauss2003, Dotter+07, VandenBerg+13}. 
Recently, \citet{PLANCK2016} have reported the reionization redshift as $7.8< z_{\rm r} <8.8$
from the Thomson scattering optical depth of the cosmic microwave background (CMB).
Based on the comparison between ages of GCs and the reionization epoch, 
most of old GCs seem to have formed under the influence of UV background radiation fields
after the cosmic reionization. 

The internal dynamics of GCs is quite distinctive from other systems with comparable luminosities 
such as dwarf spheroidal galaxies (dSphs).
GCs are very compact systems, the half-light radii ($r_{\rm h}$) of which are around 1-10 pc, 
regardless of their luminosity \citep{McConnachie12}. 
The velocity dispersions ($\sigma$) of GCs are as high as 10~km/s, and show steep dependence on luminosity ($L$) 
as $\sigma \propto L^{1/2}$ \citep[e.g.,][]{McLaughlin2000,Drinkwater,Hasegan,Forbes08}, 
which is insensitive to their radii and masses. 
These characteristic features of GCs imply that they formed in their inherent environments. 

The formation scenarios for GCs have been proposed by many authors, but still under debate. 
For instance, \cite{Kravtsov&Gnedin} have performed high-resolution cosmological simulations to explore the formation of GCs in a Milky Way (MW)-sized galaxy. 
They have found that cold metal-poor gas is supplied to the center of the galaxy by direct gas accretion along dark matter (DM) filaments
during minor mergers of smaller galaxies.  
The collisions of accreting gas spawn dense molecular clouds, which may be able to evolve to GCs. 
Although the spatial resolution of the simulations was not sufficient to resolve the internal structure of each star cluster, 
their result suggests that GCs possibly form in the cosmological context. 
The formation of giant molecular clouds can be expected also in major mergers of galaxies. 
\citet{Saitoh+09} have performed $N$-body/SPH simulations of major mergers to explore the evolution of the interstellar medium (ISM). 
As a result, they have shown that the formation of GC-sized massive star clusters is triggered 
at high dense filamentary regions compressed by shocks. Besides,
some high-resolution cosmological $N$-body simulations have revealed that the radial distribution of sub-halos 
originating from relatively rare peaks resembles the distribution of the Galactic GCs \citep{Diemand+05, Moore2006}. 
This result implies that GCs may stem from DM sub-halos, but GCs are usually observed as stellar-dominated systems. 
To reconcile this inconsistency, \cite{Saitoh+06} have shown, using a semi-cosmological hydrodynamic simulation, 
that the tidal force by a host galaxy effectively strips DM halos surrounding the star clusters. 
However, no previous work has not succeeded in accounting for
the characteristic internal properties of GCs.

As stated above, the formation of GCs is likely to be intimately related to the UV background radiation.
Many observations have shown that cosmic reionization took place around the GC formation epoch. 
For instance, \citet{ASPC_Umemura+01} 
have estimated the reionization epoch to be $6< z_{\rm r} <10$, by confronting the radiative transfer simulations
on reionization to Ly$\alpha$ absorption systems seen in high-$z$ quasar spectra.
Also, \citet{Fan+06} have estimated neutral hydrogen fractions at $z\sim 5-6$ from QSO Ly$\alpha$ absorption lines and concluded that 
reionization is almost completed by $z\gtrsim 6$. 
Besides, Gamma-Ray Bursts (GRBs) are also available to probe neutral hydrogen at high redshifts, because of their cosmological distances. 
\citet{Totani+06} have analyzed the Ly$\alpha$ damping wing in the optical afterglow spectrum of 
GRB050904 at $z=6.3$, and concluded that a large fraction of intergalactic hydrogen seems to be ionized at $z=6.3$. 	
\citet{Ouchi+10} have investigated the evolution of high-$z$ Ly$\alpha$ luminosity functions, 
and constrained the neutral hydrogen fraction in the intergalactic space as $f_{\rm HI}<0.2$ at $z=6.6$. 

UV radiation ionizes gas clouds and heats them up to $T \sim10^4$~K. 
As a result, the gravitational contraction of clouds is suppressed 
if their virial temperatures are lower than $\sim 10^4$K. 
Moreover, UV photons dissociate $\HH$ molecules that are the most important coolant at $T\lesssim 10^4$~K 
under metal-poor environments in the early Universe. 
Thus, in order for stars to form in the low-mass gas clouds exposed to UV background radiation,
the clouds should be self-shielded from a UV background \citep[][]{Tajiri&Umemura}. 
\cite{Hasegawa2009} (Hereafter HUK09) have performed spherically symmetric 
radiation hydrodynamics (RHD) simulations to explore the possibility of the star cluster formation under UV background radiation.  
As a result, they have found that the star cluster formation processes branch off into three paths 
according to the timing of the self-shielding. If the self-shielding occurs in the stage of supersonic contraction of a cloud,
it leads to the formation of very compact star clusters like GCs. 
(The details of physical processes are described in \S~\ref{sec:physical_model}). 
However, in the simulations by HUK09, only isotropic irradiation of UV was investigated.
In realistic situations, we should consider three-dimensional effects.
First, background radiation fields are usually expected to be anisotropic. 
Under an anisotropic UV background, the self-shielded regions also become anisotropic. 
Hence, the contraction of clouds is thought to proceed in a different fashion from the spherical symmetric collapse. 
Furthermore, if the density distributions in clouds are inhomogeneous, the self-shielding is subject to shadowing effects. 
In the context of the cosmic reionization, \citet{Nakamoto+01} have shown, by six-dimensional radiative transfer simulations, that 
the reionization process in an inhomogeneous media is considerably delayed compared to a homogeneous medium case 
due to the shadowing effects. 
Such inhomogeneity also increases an effective recombination rate in gas clouds, 
since the local recombination rate is proportional to the square of density \citep{Madau+1999}. 
These three-dimensional radiation hydrodynamic effects may bring significant impacts 
on the star formation in the early Universe. 

In this paper, we perform three-dimensional RHD (3D-RHD) simulations, where
the six-dimensional radiative transfer is coupled with 3D hydrodynamics, and investigate 
how the three-dimensional effects have impacts on the formation processes of star clusters 
under UV background radiation.  
This paper is organized as follows. 
In Section 2, the physical models of star cluster formation are described based on HUK09. 
Section 3 is devoted to the numerical method of the present study. 
The numerical results are presented in Section 4, where
the evolution of gas clouds exposed to external UV radiation and resultant stellar dynamics are shown. 
Also, we compare the properties of simulated star clusters to those of velocity dispersion-supported systems 
such as GCs, dSphs, and ultra compact dwarfs (UCDs). 
Finally, we discuss and conclude our results in Section 5 and 6, respectively. 
Throughout this paper, we assume a CDM cosmology neglecting the dark energy, since it is less important in the early Universe. 
We work with cosmological parameters; $\Omega_{\rm M} = 1.0$, $h=0.6777$, and $\Omega_{\rm b} = 0.1564$ \citep{Planck14}.

\section{Physical Model}\label{sec:physical_model}
To review the basic physics, we suppose a spherical gas cloud purely composed of hydrogen 
and exposed to external isotropic background radiation. 
The external UV radiation ionizes neutral hydrogen and raises the temperature up to $\sim 10^4$ K. 
Thus, if a cloud is totally ionized, the system with the mass less than 
the Jeans mass at $10^4$ K, $M_{\rm J}(T=10^4~{\rm K})$,
is hindered from collapsing owing to raised thermal pressure. 
Furthermore, the primary coolant in the temperature range of $T \sim 10^{3-4}$K is $\HH$ molecules
for the metal-poor gas with the metallicity of  $Z/\solmetal \leq 10^{-2}$ \citep{Susa&Umemura2004}. 
The Lyman-Werner band UV radiation dissociates $\HH$ molecules, so that stars cannot be born in the cloud. 
Therefore, in order for star clusters to form, 
the cloud should be self-shielded from ionizing and dissociating UV radiation.
\citet{Tajiri&Umemura} have studied the condition for the self-shielding by solving radiative transfer in the spherical symmetric geometry. 
Assuming power-law UV background radiation intensity $I_\nu = 10^{-21} \times I_{21}\left(\nu/\nu_{\rm L}\right)^{-1}$ erg $\per{cm}{2}$ $\per{s}{1}$ $\per{str}{1}$ $\per{Hz}{1}$, where $\nu_L$ indicates the Lyman limit frequency and $I_{21}$ is the intensity at the Lyman limit frequency in units of $10^{-21}~{\rm erg}~\per{cm}{2}~\per{s}{1}~\per{str}{1}~\per{Hz}{1}$, they have shown that the critical number density $n_{\rm crit}$ required for being shielded against the ionizing background radiation is given by
\begin{equation}
	\label{eq:n_crit}
	n_{\rm crit} = 1.40\times10^{-2}{\per{cm}{3}}\left(\frac{M}{10^8\solmass}\right)^{-1/5}I_{21}^{3/5}, 
\end{equation}
or the corresponding critical radius is 
\begin{equation}
	\label{eq:r_crit}
	r_{\rm crit} = 4.10{\rm kpc}\left(\frac{M}{10^8\solmass}\right)^{2/5}I_{21}^{-1/5}, 
\end{equation}
where $M$ is the total mass of the cloud. 
The self-shielded regions inside the shielding radius ($r_{\rm shield}$) can gravitationally contract, 
even if the mass $M(<r_{\rm shield})$ is less than $M_{\rm J}$($10^4$K). 
Also, if the interior mass is massive enough to produce $\HH$ molecules ($T_{\rm vir}>10^3$K)
and also the dissociating UV radiation is shielded \citep{Draine&Bertoldi1996},
stars can form in the cloud. 
Thus, the star cluster formation under a UV background is regulated by the self-shielding condition. 

\begin{figure}
	\begin{center}
		\includegraphics[width=85mm,clip]{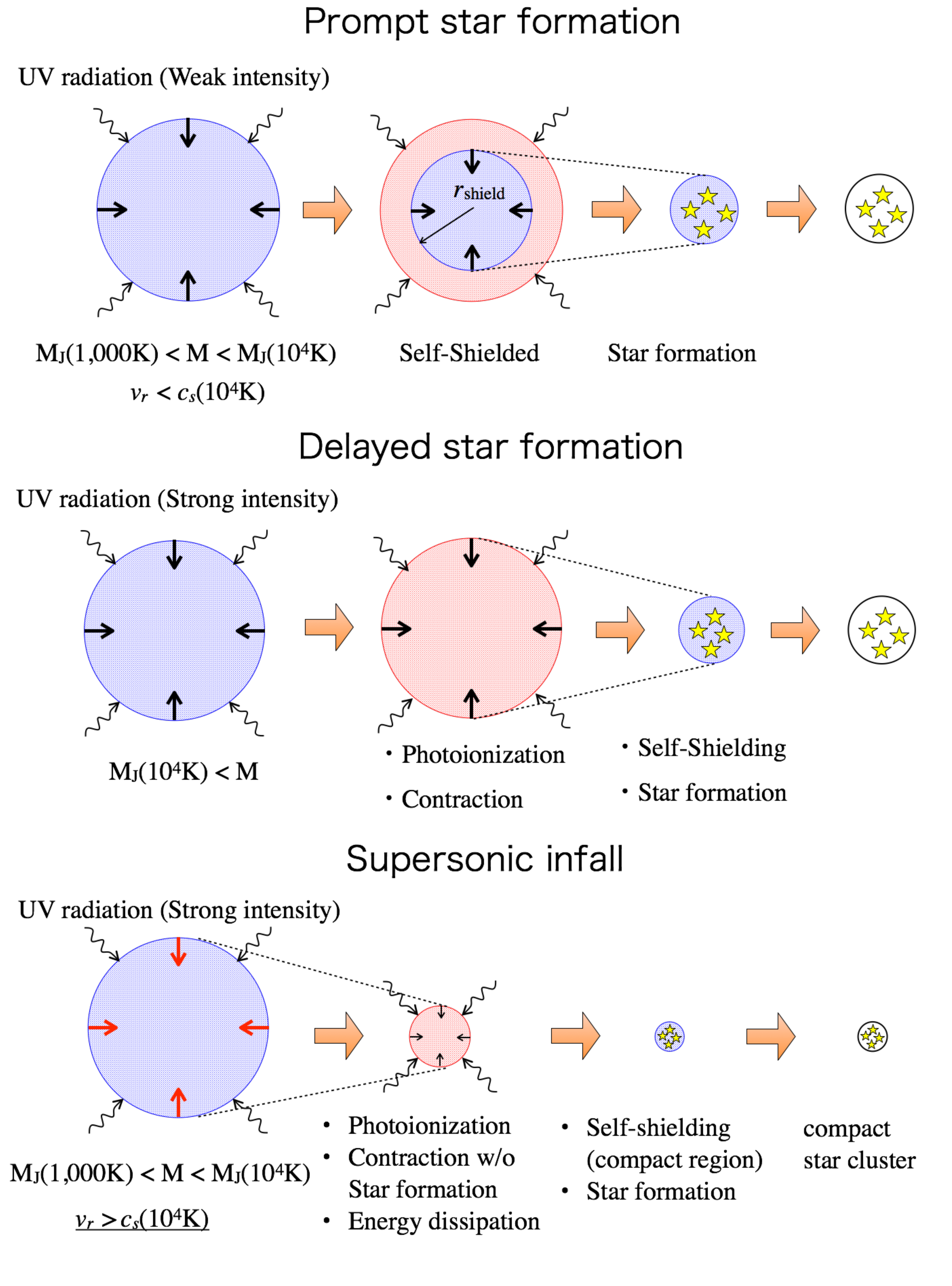}
%
%
	\end{center}
	\caption{Schematic views for the formation scenarios of star clusters proposed by HUK09. 
		The upper, middle, and bottom panels represent the ``prompt star formation'', ``delayed star formation'', and ``supersonic infall star formation'' scenarios, respectively. 
		In each panel, the red shaded region represents the photo-ionized gas, while the blue shaded denotes the self-shielded neutral regions. 
		Red arrows denote the infall velocity exceeding the sound speed of photo-ionized gas ($\sim$ 10 km/s).  }
	\label{fig:concept_Hasegawa09}
\end{figure}

In the followings, we briefly describe three branches of the star cluster formation  
regulated by the self-shielding, based on HUK09. 
The schematic views for the formation scenarios of star clusters are presented in Fig.~\ref{fig:concept_Hasegawa09}. 

\subsection{Prompt Star Formation}
When the cloud mass $M$ is in the range of $M_{\rm J}(10^3 {\rm K})<M<M_{\rm J}(10^4 \rm {K})$ and incident UV intensity is relatively weak (i.e., the cloud radius is smaller than $r_{\rm crit}$), the self-shielded region promptly forms inside the cloud. 
As a result, the star formation is initiated in the self-shielded region, while the outer region is conversely evaporated by photo-heating. 
This branch is called ``prompt star formation'', which is the basic mechanism 
for the formation of low-mass galaxies during cosmic reionization \citep[e.g.,][]{Susa&Umemura2004}. 

In this case, stars can begin to form at an early stage of the contraction. 
Hence, the gas is effectively converted to stars without dissipating a large amount of kinetic energy. 
As a result, diffuse stellar systems tend to form. 
HUK09 have shown that the resultant star clusters mimic dSphs on the $\sigma$-$L$ plane.  

\subsection{Delayed Star Formation}
When the cloud mass exceeds the Jeans mass of photo-ionized gas, i.e., $M>M_{\rm J}(10^4 \rm {K})$, 
the cloud can collapse, even if it is totally ionized by strong UV radiation. 
Although the cloud can keep shrinking, stars are never born before the cloud is self-shielded. 
Hence, the star formation tends to be delayed compared to no or weak UV cases. 
This mechanism is called ``delayed star formation''. 

In this case, the gas cloud can collapse without mass-loss, but some amount of the kinetic energy is dissipated. 
As a result, the systems formed via the ``delayed star formation'' exhibit relatively high velocity dispersions. 

\subsection{Supersonic Infall Star Formation}
Finally, we argue the case that the cloud mass is in the range of $M_{\rm J}(10^3 {\rm K})<M<M_{\rm J}(10^4 \rm {K})$, 
and strong UV is irradiated to the cloud. 
If the cloud radius is larger than the critical radius $r_{\rm crit}$, the bulk of the cloud are ionized.
In this case, the cloud cannot collapse due to the thermal pressure of ionized gas. 
However, HUK09 have shown that the cloud can collapse if the cloud contracts
with infall velocity exceeding the sound speed of ionized gas. 
The contraction continues until the self-shielding effects work, and eventually stars can form in the self-shielded regions. 
This branch is called ``supersonic infall star formation''. 
In this case, the star-forming regions become very compact, 
and the infall velocity is strongly decelerated due to the thermal pressure. 
As a result, the star clusters formed via the ``supersonic infall'' can be as compact as GCs.  

Also, HUK09 have argued the probability of such ``supersonic infall'' in the context 
of the CDM cosmology. They have found that the ``supersonic infall'' under a strong
UV background can occur at redshift $z \gtrsim 15$ in density peaks higher
than the standard deviation of density fluctuations.

\section{method}\label{sec:method}
We perform 3D-RHD simulations to investigate the multi-dimensional effects in the formation of star clusters under UV background radiation. 
In the simulations, we consistently solve three-dimensional hydrodynamics, non-equilibrium chemistry, the transfer of UV photons, and the gravitational force. 

\subsection{Three-Dimensional Hydrodynamics}
We solve hydrodynamics by Smoothed Particle Hydrodynamics (SPH) method, utilizing the code based on \citet{HUS09} and \citet{START} that have been developed to solve hydrodynamics coupled with the radiative transfer of UV photons. 

We set the smoothing length $h$ so that the mass enclosed in the sphere with the radius $h$ is held constant. 
At every timestep, the smoothing length of each SPH particle is iteratively determined to satisfy the above condition \citep{Price&Monaghan07}. 
As for the gravitational force calculation, we adopt the Barnes-Hut Tree-algorithm to reduce the numerical cost \citep{Barnes&Hut86}, setting the opening angle to be $\theta = 0.5$. 

\subsection{Chemical Reactions and Three-Dimensional Ray-Tracing}
In the simulations, we solve primordial chemical networks regarding six species, i.e., $\electron$, $\Hp$, $\rm H$, $\Hm$, $\HH$, and $\HHp$. 
We neglect cooling processes by metals, because the major coolant in the temperature range 
of $T \sim 10^{3-4}$K is $\HH$ as long as the metallicity is $Z/\solmetal \leq 10^{-2}$ \citep{Susa&Umemura2004}. 
This is a reasonable approximation for the present study, since we focus on the star formation under metal-poor environments in the early Universe. 

For the HI photo-ionization process, we assume the on-the-spot approximation \citep{Spitzer78}, 
in which ionized photons emitted from the recombination to the ground state of hydrogen are assumed to be absorbed on the spot.   
In this case, the solution of the radiative transfer equation is simply given by $I_\nu(\tau_\nu)=I_0\exp \left(-\tau_\nu \right)$, where $I_0$ and $\tau_\nu$ are respectively the intensity of a UV source and the optical depth at a frequency $\nu$. 
We integrate the optical depth between a UV source and a target SPH particle by the same way as RSPH method \citep{RSPH}, and evaluate the photo-ionization and photo-heating rates. 
To calculate $\HH$ photo-dissociation rates, we adopt the self-shielding function derived by \citet{Draine&Bertoldi1996}. 
Using the self-shielding function, the photo-dissociating radiation flux is given by 
\begin{equation}
F = F_0f_{\rm sh}(N_{\HH,14}), 
\end{equation}
where $F_{0}$ is the flux without the self-shielding, $N_{\HH,14}$ is the $\HH$ column density normalized by $10^{14}\;\rm cm^{-2}$, and $f_{\rm sh}$ is 
\begin{equation}
f_{\rm sh}(x)= \left \{
\begin{array}{l}
1,\;\;\;\;\;\;\;\;(x\le1)\\
x^{-\frac{3}{4}}.\;\;\;\;(x>1)
\end{array}
\right.
\end{equation}
The $\HH$ column density is also evaluated by RSPH method. 
We should mention that the dependence of the $\HH$ self-shielding on the column density is weaker than that of the $\rm HI$ shielding, and hence photo-dissociating photons are more permeable than ionizing photons, leading to deeper suppression on the star formation.  
The photo-detachment of $\Hm$ and the photo-dissociation of $\HHp$ are also considered with the optically-thin approximation because of their very small fractions.   
The cross-sections for these processes are taken from \citet{Tegmark+97} and \citet{Stancil94}.

\begin{table*}
\begin{center}
\caption{Numerical parameters in all runs}
\label{table:parameter}
\begin{tabular}{ccccccc}
\hline \hline
$z_{\rm c}$ & 
radiation field &
formation & 
$z_{\rm UV}$ &
$M_{\rm ini} $ &
$ \dot{N}_{\rm ion}/\dot{N}_{\rm crit} $ &
$t_{\rm rise}$ \\
& & & & $[10^6\solmass]$ &  & [yr]
\\
\hline
    6 & one-sided/isotropic & supersonic & 6.8 & 2.5 & 10 & instant \\
    6 & one-sided/isotropic & supersonic & 6.9 & 5.0 & 10 & instant \\
    9 & one-sided/isotropic & supersonic& 10.3 & 2.5 & 10 & instant  \\
    9 & one-sided/isotropic & supersonic& 10.5 & 5.0 & 10  & instant \\
    12 & one-sided/isotropic & supersonic& 13.8 & 2.5 & 10 & instant \\
    12 & one-sided/isotropic & supersonic& 14.0 & 5.0 & 10 & instant \\ 
    9 & one-sided & supersonic& 10.5 & 5.0 & 10  & $10^6$\\
    9 & one-sided & supersonic& 10.5 & 5.0 & 10  & $10^7$ \\
    9 & one-sided & supersonic& 10.5 & 5.0 & 10  & $10^8$ \\
    9 & one-sided & supersonic & 10.5 & 10.0 & 10  & instant \\
    12 & one-sided & supersonic & 14.0 & 10.0 & 10 & instant  \\
    6 & one-sided & prompt & 8 & 2.5 & 0.1 & instant \\
    6 & one-sided & prompt & 8 & 5.0 & 0.1  & instant \\
    9 & one-sided & prompt & 8 & 1.0 & 0.1  & instant \\
    9 & one-sided & prompt & 12 & 2.5 & 0.1  & instant \\
    9 & one-sided & prompt & 12 & 5.0 & 0.1  & instant \\
    9 & one-sided & prompt & 12 & 10.0 & 0.1 & instant \\
    12 & one-sided & prompt & 15.9 & 1.0 & 0.1  & instant \\
    12 & one-sided & prompt & 15.9 & 2.5 & 0.1  & instant \\
    12 & one-sided & prompt & 15.9 & 5.0 & 0.1  & instant \\ \hline
\end{tabular}
\end{center}
\end{table*}

\subsection{Setup}\label{sec:setup}
In each run, we consider a low-mass gas cloud with the initial baryonic mass 
of $10^6\solmass  \leq$ $M$ $\leq10^7\solmass$ in a DM halo that collapses at a redshift of $6 \leq z_{\rm c} \leq 12$. 
As for the initial chemical composition, we refer to the cosmological pre-reionization values derived by \citet{Galli&Palla98}. 
We assume the initial temperature to be $T=100$ K referring to \citet{Iliev06, Iliev09}. 
We start each simulation from the stage when a cloud reaches the maximum expansion. 
The maximum expansion redshift $z_{\rm max}$ is related to the collapse redshift $z_{\rm c}$ as $(1+z_{\rm max}) = 2^{2/3}(1+z_{\rm c})$. 
The maximum expansion radius is given by
\begin{equation}
	\label{eq:Rmax}
		r_{\rm max}= \left(\frac{4M}{3\pi^3 \rho_{\rm c0}}\right)^{1/3}(1+z_{\rm max})^{-1}, 
\end{equation}
where $\rho_{\rm c0}$ is the cosmic critical density at the present-day. 
In the mass range we consider, the mass resolution is set to be roughly the same. In the present simulations, we use $2^{15-18}$ SPH particles
according to the cloud mass. 
Thus, the SPH particle mass is $m_{\rm SPH}\approx  40\solmass$ and 
the effective mass resolution of hydrodynamics is $\approx 4000 \solmass$ \citep{Bate&Burkert1997, Bate+2003}. 
This mass resolution allows us to pursue the local Jeans instability of primordial gas up to $\sim 10^5$ cm$^{-3}$ \citep[e.g.,][]{Omukai+05}. 
The number of DM particles is set to be the same as that of SPH particles, and the DM particle mass corresponds to $m_{\rm DM}\approx m_{\rm SPH}(\Omega_{\rm M}-\Omega_{\rm b})/\Omega_{\rm b}$. 
The initial density profiles of gas and dark matter are assumed to be in the form of $\delta(r) = \delta_0\sin(\lambda r)/(\lambda r)$ \citep{Kitayama+01}, where $\lambda$ is defined as $\lambda r_{\rm max} = \pi$. 
We set $\delta_0$ so that the averaged overdensity within $r_{\rm max}$ is equivalent to 4.55, which is the value derived from the analytic spherical collapse model.  
We also add inhomogeneity to the cloud. 
We generate random-Gaussian density fields, where the power spectrum of density fluctuations obeys $P(k) =A_{\rm amp} k^{p}$
for the wave number $k$. 
We use the same method as \citet{Braun+1988} to generate the random-Gaussian density fields. 
We set the power-law index $p = -3$ to mimic scale-free density fluctuations of cold dark matter. 
The amplitude $A_{\rm amp}$ is related to the initial clumping factor $C = \langle n^2 \rangle/\langle n\rangle^2$. 
In SPH simulations, the clumping factor can be evaluated by the simple formula 
\begin{equation}
C = \frac{\sum_i m_i \rho_i^{-1}\sum_j m_j \rho_j}{\left(\sum_k m_k \right)^2}, 
\end{equation}
where $\rho_i$ and $m_i$ are the density and mass of the $i$-th particle, respectively \citep{Springel&Hernquist2003}. 
The initial clumping factor is $C=1.7$ in all of the runs in this paper.  

Also, we specify the epoch of the irradiation of external UV radiation $z_{\rm UV}$, 
which are greater than $z=6$, according to the reionization epoch suggested by observations.  
As for the intensity of UV background, we define the critical number of incident ionizing photons per unit time $\dot{N}_{\rm crit}$,
which is required to ionize the entire volume of gas cloud $V_{\rm UV,in}$.
$\dot{N}_{\rm crit}$ is given by a following formula \citep{Madau+1999}; 
\begin{equation}
	\label{eq4}
	\dot{N}_{\rm crit} =  \langle n \rangle^2 \alpha_{\rm B}(T)CV_{\rm UV,in}, 
\end{equation}
where $\alpha_{\rm B}(T)$ is the case B recombination coefficient of hydrogen, $\alpha_B =2.59\times10^{-13} \per{cm}{3}\per{s}{1}$ at $T = 10^4$ K. 
We specify the number of incident UV photons $\dot{N}_{\rm ion}$ in units of $\dot{N}_{\rm crit}$. 
In our simulations, $\dot{N}_{\rm ion}/\dot{N}_{\rm crit} = 10$ or $\dot{N}_{\rm ion}/\dot{N}_{\rm crit} = 0.1$ 
are assumed so that we can explore the dependence on the strength of UV background radiation. 
Also, in some models, we consider the finite time over which 
the luminosity rises up to $\dot{N}_{\rm ion}$. Here, we simply assume the linear rising of the luminosity, i.e., 
$\dot{N}(t) = \dot{N}_{\rm ion}/t_{\rm rise} \times t$, where $t_{\rm rise}$ denotes the rising time,
and after $t_{\rm rise}$ the luminosity is set to be constant, $\dot{N}(t) = \dot{N}_{\rm ion}$. 
In this paper, we examine three cases of $t_{\rm rise}$ as 1 Myr, 10 Myr, and 100 Myr. 
To investigate the effect of the anisotropy of radiation fields, we adopt two extreme cases; the one-sided or isotropic background radiation. 
In each run with a one-sided radiation field, we place only one ionizing source. 
The direction towards the source is referred to as the $x$-direction. 
On the other hand, in each run with an isotropic radiation field, we isotropically distribute 18 sources around a gas cloud. 
Note that the luminosity per one source in the one-sided radiation field case is 18 times higher than that in the isotropic radiation field case if $\dot{N}_{\rm ion}/\dot{N}_{\rm crit}$ is the same. 
As a UV spectrum, we assume the black body type with the effective temperature of $T_{\rm eff} =10^5$K, since young massive stars are generally thought to be dominant ionizing sources during the reionization epoch. 

\subsection{Star Formation and Dynamical Evolution}
We incorporate the star formation in self-shielded, cooled regions. 
Here the following star formation criteria are employed; 
(1) $\nabla \cdot \vec{v}<0$, (2) $y_{\HH} \geqslant 5\times10^{-4}$, and (3) $T \leqslant 5000 \rm K$, where $\vec{v}$, $y_{\HH}$, and $T$ are the local velocity, the $\HH$ fraction, and the gas temperature, respectively. 
In particular, the condition (2) is never satisfied unless the gas is shielded against ionizing and dissociating photons.  
Thus, this is an essential condition that regulates the star formation. 

We consider the timescale in which a gas particle is converted into a collisionless star particle. 
At every timestep, we search gas particles satisfying the above criteria, and convert them to collisionless stellar particles stochastically as follows; 
The star forming timescale is expected to be controlled by the local free-fall timescale $t_{\rm ff} = \sqrt{3\pi/32G\rho_{\rm gas}}$, where $\rho_{\rm gas}$ is the local gas density. 
Using the free-fall time $t_{\rm ff}$, we describe the star formation rate as 
\begin{equation}
\label{ }
	\diff{}{\rho_\ast}{t} = c_\ast \frac{\rho_{\rm gas}}{t_{\rm ff}}, 
\end{equation}
where $\rho_\ast$ is the local stellar density, and $c_\ast$ is the dimensionless parameter to control star formation efficiency. 
We can determine the probability $p_\ast$ that SPH particles of $m_{\rm gas}=N_{\rm neighbor}m_{\rm SPH}$ are converted to 
stellar particles of $m_\ast=\alpha_\ast \times m_{\rm gas}$($0< \alpha_\ast <1$) during the time interval $\Delta t$: 
\begin{equation}
\label{ }
	p_\ast = \alpha_\ast^{-1}\left[1-\exp \left(-c_\ast \frac{\Delta t}{t_{\rm ff}}\right)\right]. 
\end{equation}
In this paper, we assume $\alpha_\ast = 0.3$ \citep{Okamoto+2003} and $c_\ast=1.0$. 
Here, $c_\ast$ is an artificial parameter to control the star formation efficiency (SFE) in the numerical simulations. 
However, as shown by \citet{Susa&Umemura2004}, the final stellar mass fractions are expected 
to be almost independent of $c_\ast$, since the SFE is essentially regulated by the self-shielding. 
We continue each run until stars formed in the cloud are settled in the quasi-steady state.  
The parameter sets in this work are summarized in Table \ref{table:parameter}.

\section{Results}
\subsection{Evolution of Gas Clouds}
\subsubsection{Supersonic Infall 
}\label{SS}
\begin{figure*}
	\begin{center}
		\includegraphics[width=14cm,clip]{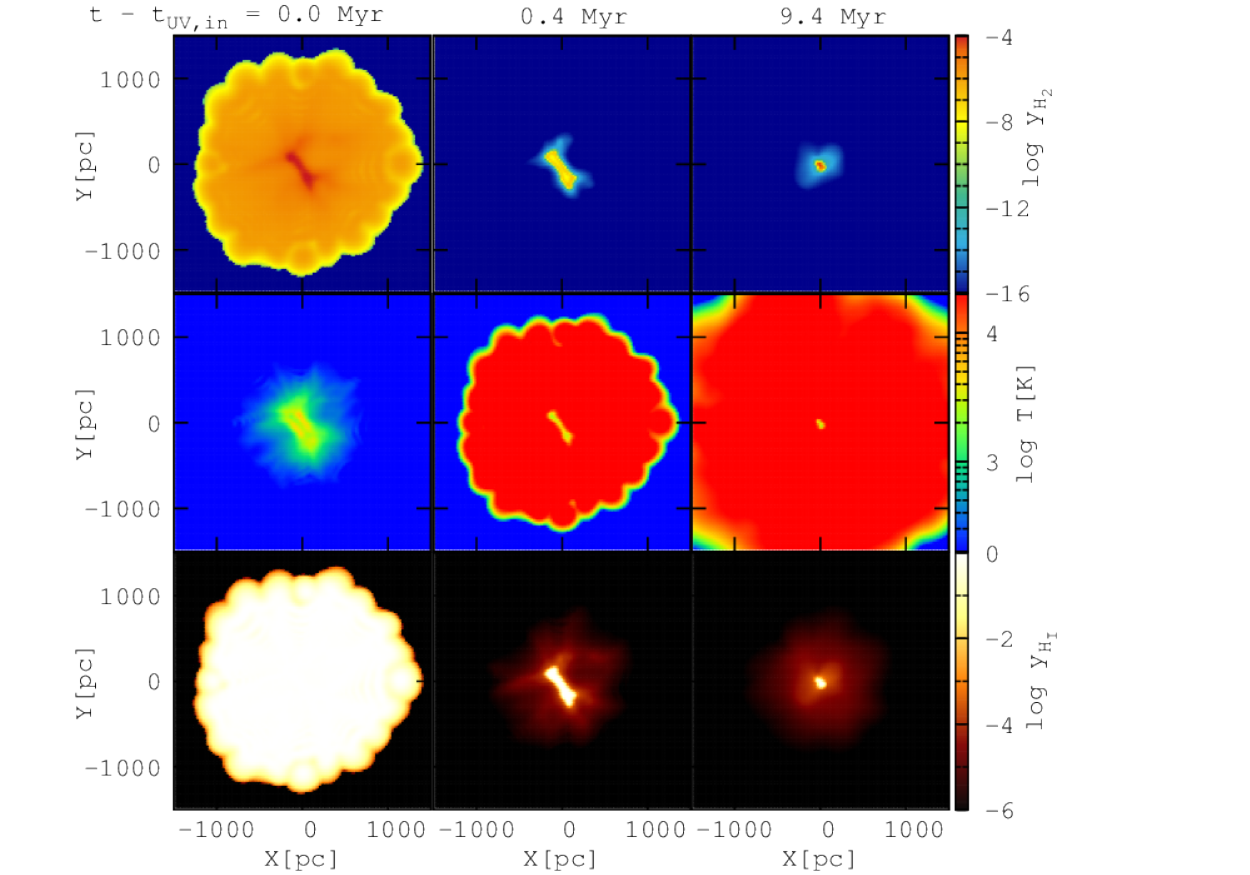}
		\caption{Time evolution of $\HH$ fraction (top panels),  temperature (middle panels), and 
		$\rm HI$ fraction distributions (bottom panels) on the $x$~-~$y$ plane in the case of the isotropic background radiation. 
		The parameters are $M_{\rm ini}=5\times 10^6\solmass$, $z_c = 9$, $z_{\rm UV}=10.5$, and $\dot{N}_{\rm ion}/\dot{N}_{\rm crit} = 10$. 
		From left to right, the distributions are shown at the UV irradiation epoch, 0.4~Myr later, and 9.4~Myr later, respectively.}			
	\label{fig:color-5M6-zcoll9-iso}
	\end{center}
\end{figure*}
\begin{figure*}
	\begin{center}
		\includegraphics[width=14cm,clip]{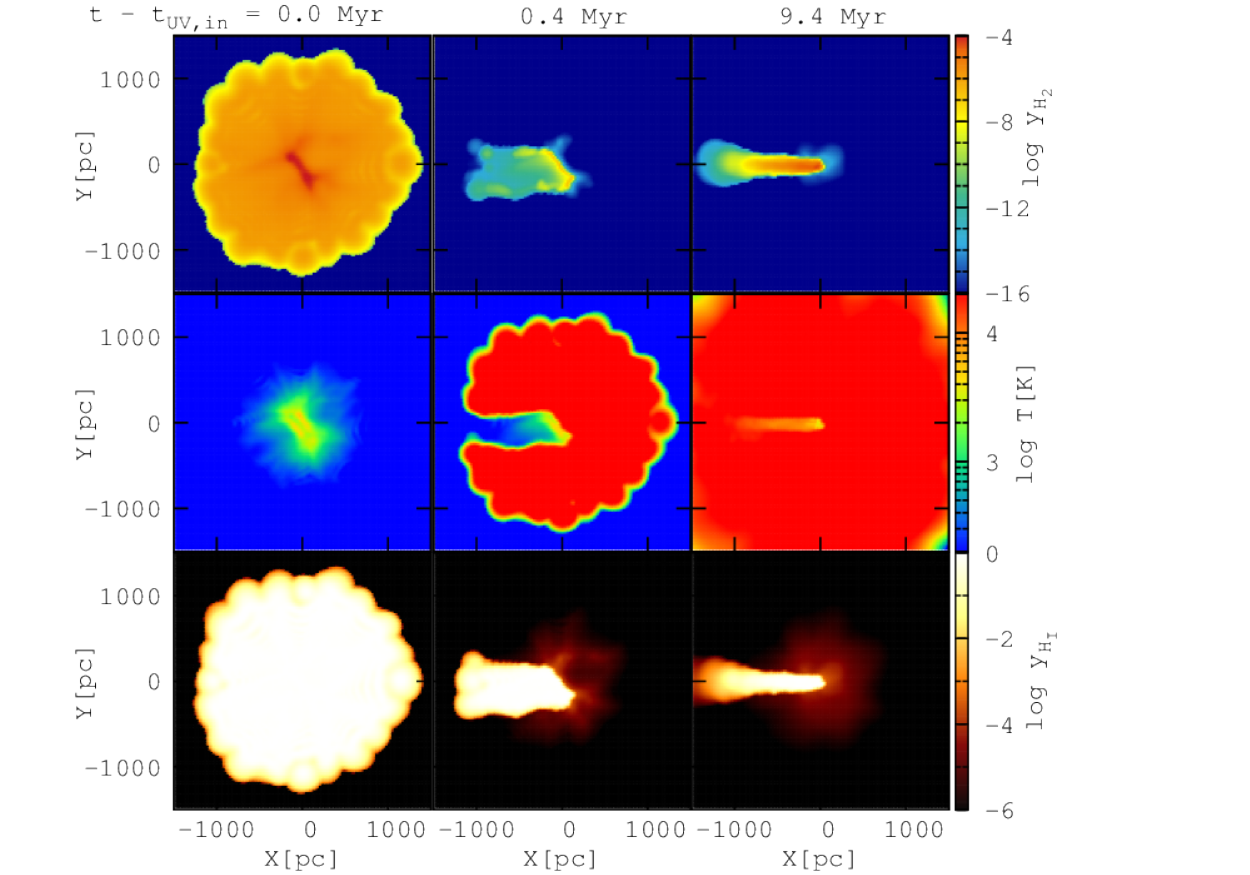}
		\caption{Same as Fig. \ref{fig:color-5M6-zcoll9-iso}, but for the case of the one-sided background radiation.}			
	\label{fig:color-5M6-zcoll9-1src}
	\end{center}
\end{figure*}
\begin{figure*}
	\begin{center}
		\includegraphics[width=14cm]{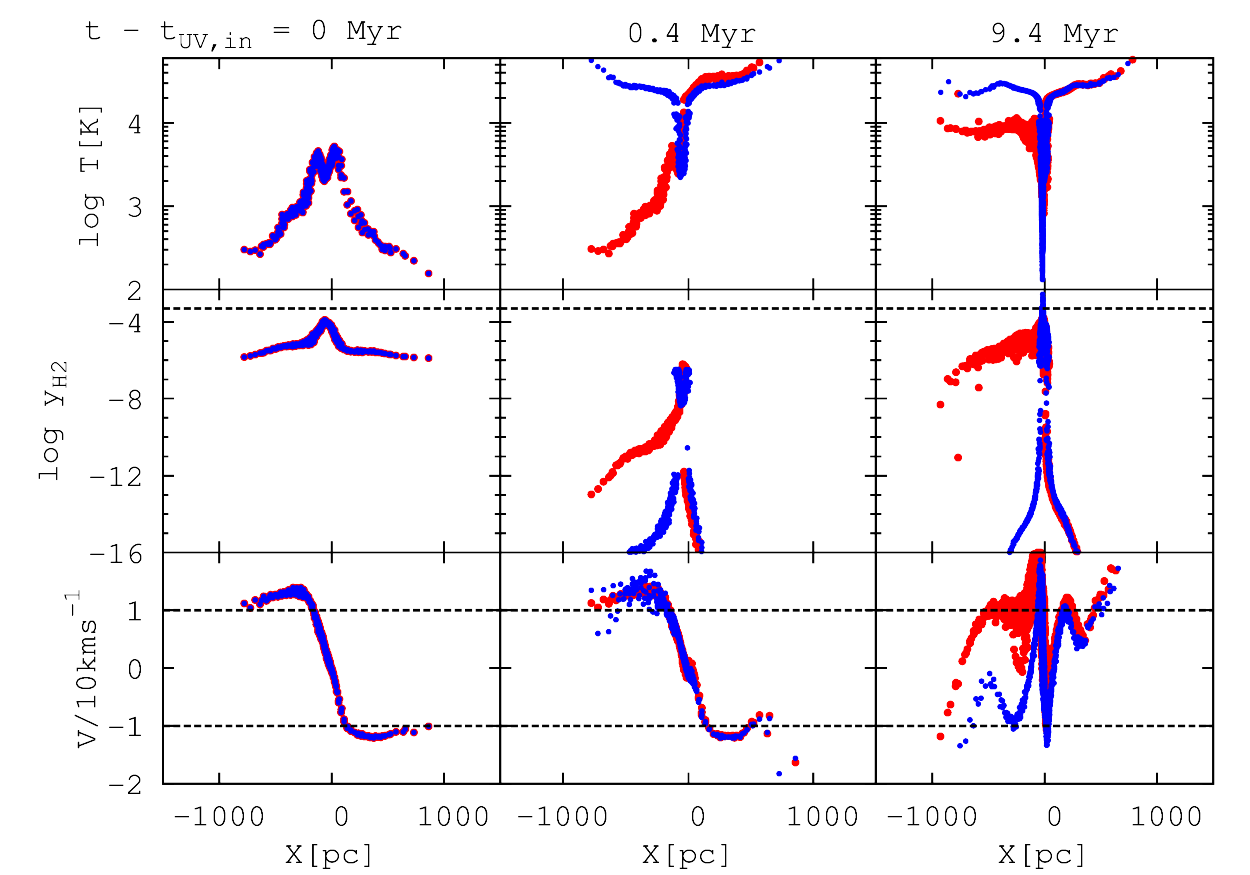}
		\caption{Temperature $T$ (top), $\HH$ fraction (middle), 
		and velocity (bottom) along the $x$-axis through the center of the cloud. 
		The red and blue points respectively indicate the quantities in the one-sided and isotropic background radiation cases.  
		A dashed line in the middle panel at each epoch indicates the star formation criteria of $y_{\HH} = 5\times 10^{-4}$. 
Two dashed lines in the bottom panel correspond to the infall velocity of 10~km/s, which is roughly the sound speed of photo-ionized gas. 
		The parameters ($M_{\rm ini}$, $z_{\rm c}$, $z_{\rm UV}$ and $\dot{N}_{\rm ion}/\dot{N}_{\rm crit}$) and the time sequence are the same as those in Fig.~\ref{fig:color-5M6-zcoll9-iso}. 
		 }
	\label{fig:1D-profile-5M6-zcoll9}
	\end{center}
\end{figure*}

First, we see the evolution of a cloud infalling with supersonic speed under a strong UV background, and
investigate the dependence on the anisotropy of UV background radiation. 
The cloud evolution under an isotropic UV background is shown in Fig.~\ref{fig:color-5M6-zcoll9-iso}
for a run with $M_{\rm ini}=5\times 10^6\solmass$, $z_{\rm c}=9$, $z_{\rm UV}=10.5$, and $\dot{N}_{\rm ion}/\dot{N}_{\rm crit}=10$,
where  the distributions of $\HH$ fraction, temperature, and $\rm HI$ fraction are presented. 
As we can see, although the outer envelope is evaporated,  $\HH$ molecules in the central region are produced abundantly, which allow the star formation. 
On the other hand, Fig. \ref{fig:color-5M6-zcoll9-1src} shows the evolution of a gas cloud exposed to one-sided background radiation. 
In this case, the shaded regions appear on the opposite side of the ionizing source owing to the shadowing effect. 
Accordingly, $\HH$ molecules form in an elongated region. 

To understand the evolution in more detail, we show the temperature, $\HH$ fraction, and velocity profiles 
along the $x$-axis through the center of the cloud in Fig.~\ref{fig:1D-profile-5M6-zcoll9}, where 
the blue points show the results for isotropic UV background radiation and
the red points for one-sided UV radiation. 
At the initial epoch, the infall velocity exceeds 10~km/s, which roughly corresponds to the sound speed of the photo-ionized gas.
Owing to such a high infall velocity, the ionized gas with $\sim10^4$~K can keep contracting even after the UV irradiation. 
As shown in Fig.~\ref{fig:1D-profile-5M6-zcoll9},
the temperature and $\HH$ molecule distributions are obviously different between isotropic and one-sided background radiation.
For isotropic background, only the central region is self-shielded from ionizing radiation and the temperature can keep below $10^4$K at 0.4Myr,
and cools down to $<10^3$K due to $\HH$ molecules at 9.4Myr.
In the case of one-sided background radiation, the temperature is below $10^4$K on the opposite side of the UV source at 0.4Myr,
whereas at 9.4Myr the temperature is raised by the weak ionization 
due to the reduction of shadowing effect caused by the shrink of the core. 
Then, the weak ionization enhances the $\HH$ molecule formation, since free electrons are the catalyst of  $\HH$ formation through the $\Hm$ process.
Also, the velocity profile for one-sided background radiation shows the weak expansion of cloud envelope, which leads to the mass loss
from the system. 
The impacts of such three-dimensional shadowing effects on the star formation 
are argued in \S\ref{section:StarFormation} later on.

\subsubsection{Prompt Star Formation}\label{PS}

\begin{figure*}
	\begin{center}
		\includegraphics[width=14cm,clip]{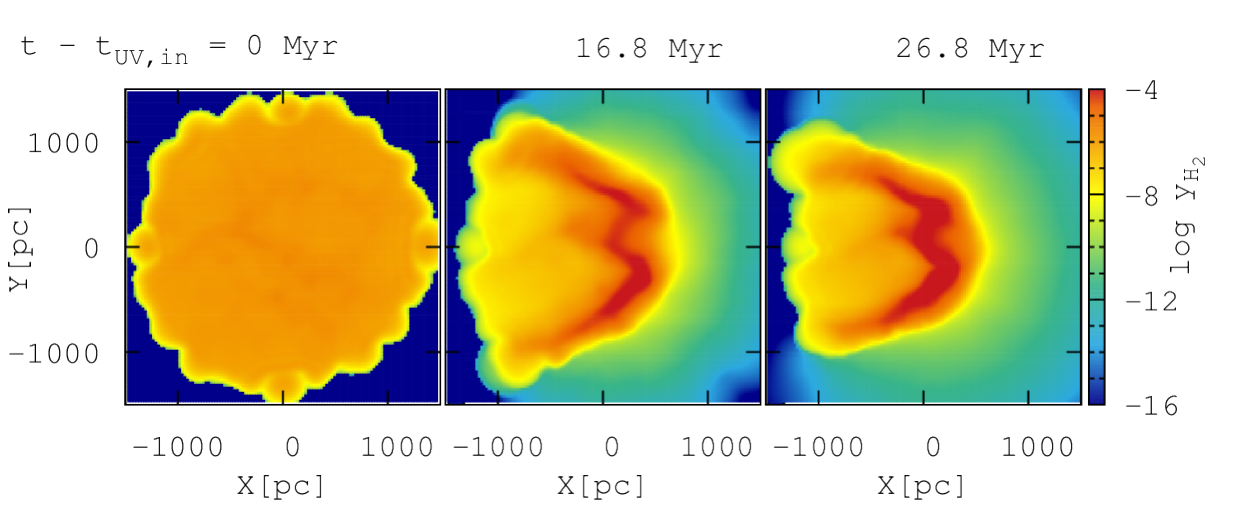}
		\caption{Time evolution of $\HH$ distributions in the run with $M_{\rm ini}=5\times 10^6\solmass$, $z_c = 9$, $z_{\rm UV}=12$, and $\dot{N}_{\rm ion}/\dot{N}_{\rm crit} = 0.1$ on the $x$~-~$y$ plane. 
			The left, center, and right panels correspond to the distributions at 0~Myr, 16.8~Myr, and 26.8~Myr after the UV irradiation, respectively. }
		\label{fig:color-H2-5M6-zcoll9-prompt}
	\end{center}
\end{figure*}
\begin{figure*}
	\begin{center}	
	\includegraphics[width=14cm]{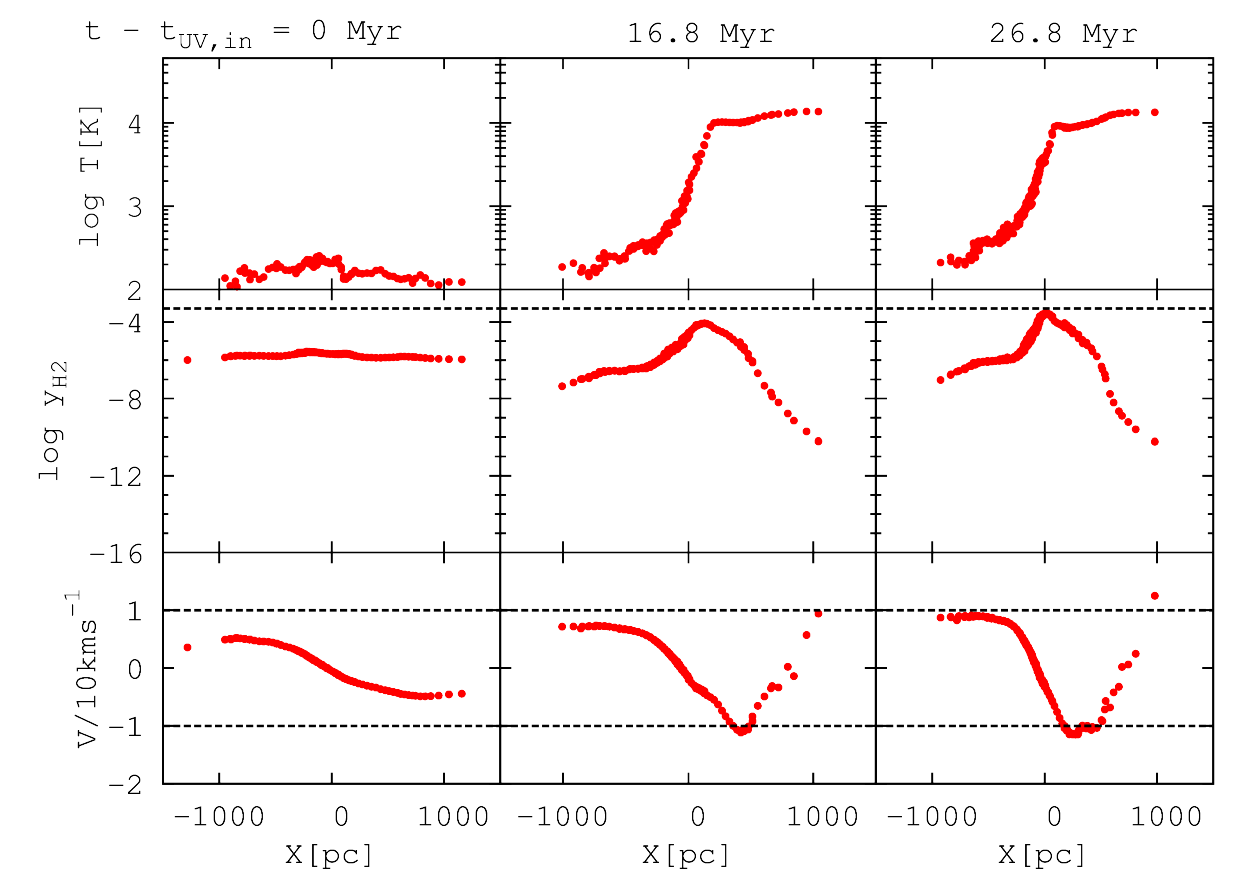}
	\caption{Same as Fig. \ref{fig:1D-profile-5M6-zcoll9}, 
		but for the cloud shown in Fig.~\ref{fig:color-H2-5M6-zcoll9-prompt}. 
		From left to right, the three panels respectively represent a cloud at the moment of 0~Myr, 16.8~Myr, and 26.8~Myr after the UV irradiation.}
	\label{fig:1D-profile-5M6-zcoll9_prompt}
	\end{center}
\end{figure*}

Here, we consider the evolution of a cloud infalling with subsonic speed under weak UV background radiation.
Fig.~\ref{fig:color-H2-5M6-zcoll9-prompt} shows the time evolution of the $\HH$ distributions in the run 
with $M_{\rm ini}=5\times 10^6\solmass$, $z_{\rm c}=9$, $z_{\rm UV}=12$, and $\dot{N}_{\rm ion}/\dot{N}_{\rm crit}=0.1$.
Fig.~\ref{fig:1D-profile-5M6-zcoll9_prompt} presents the temperature, $\HH$ fraction, 
and velocity profiles along the $x$-axis through the cloud center. 
Since this cloud is irradiated by external radiation at an earlier phase of its contraction compared to the run shown in \S~\ref{SS}, 
the infall velocity is lower than the sound speed of photo-ionized gas at the moment of the irradiation. 
Hence, the ionized gas inevitably evaporates, and the self-shielded regions can collapse.
One of the notable phenomena in this case is the positive feedback by ionizing photons. 
As shown in Fig.~\ref{fig:1D-profile-5M6-zcoll9_prompt}, $\HH$ molecules are efficiently formed 
around the ionization front. 
This is caused by the increase of free electrons that act as the catalyst for the $\HH$ formation. 
The enhanced $\HH$ formation induces the formation of stars at the positive feedback region \citep{Ricotti02}, 
and simultaneously protects the central region of the cloud from photo-dissociating radiation \citep{Susa+09,HUS09}.
As a result, the ``prompt star formation'' proceeds in the cloud, as proposed by HUK09.

\begin{figure*}
	\begin{center}

	\begin{minipage}{0.33\hsize}
		\begin{center}
			\includegraphics[width=60mm,clip]{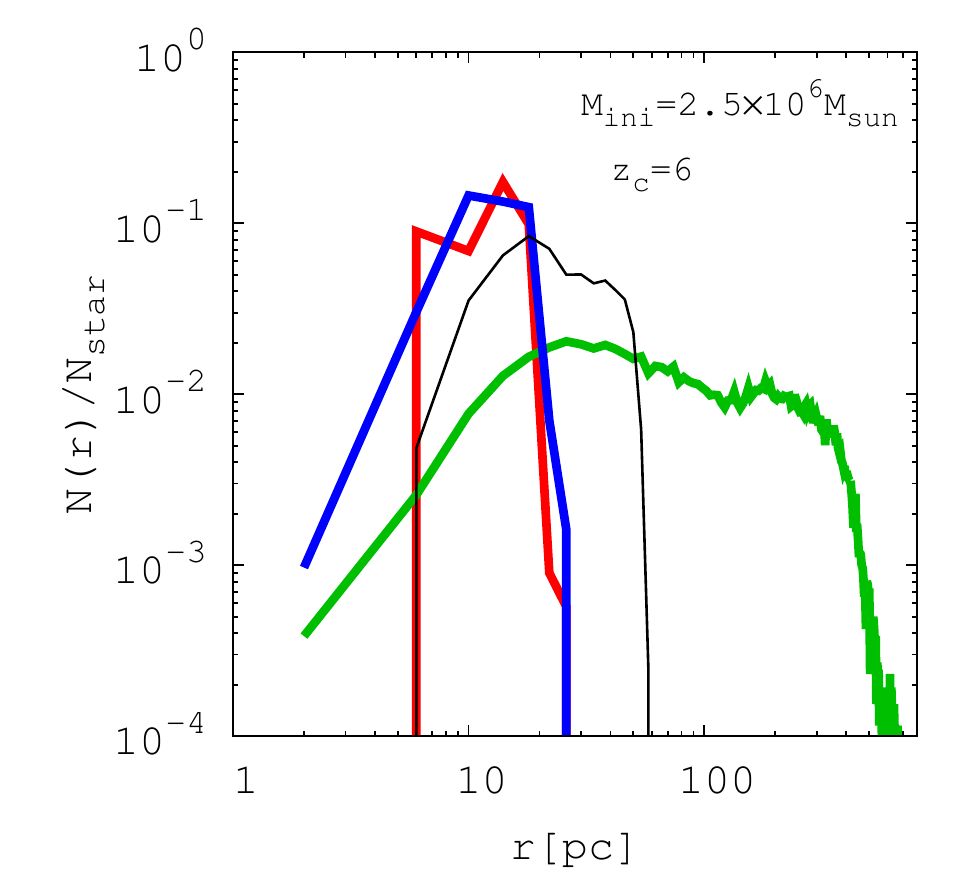}
		\end{center}
	\end{minipage}
	\begin{minipage}{0.33\hsize}
		\begin{center}
			\includegraphics[width=60mm,clip]{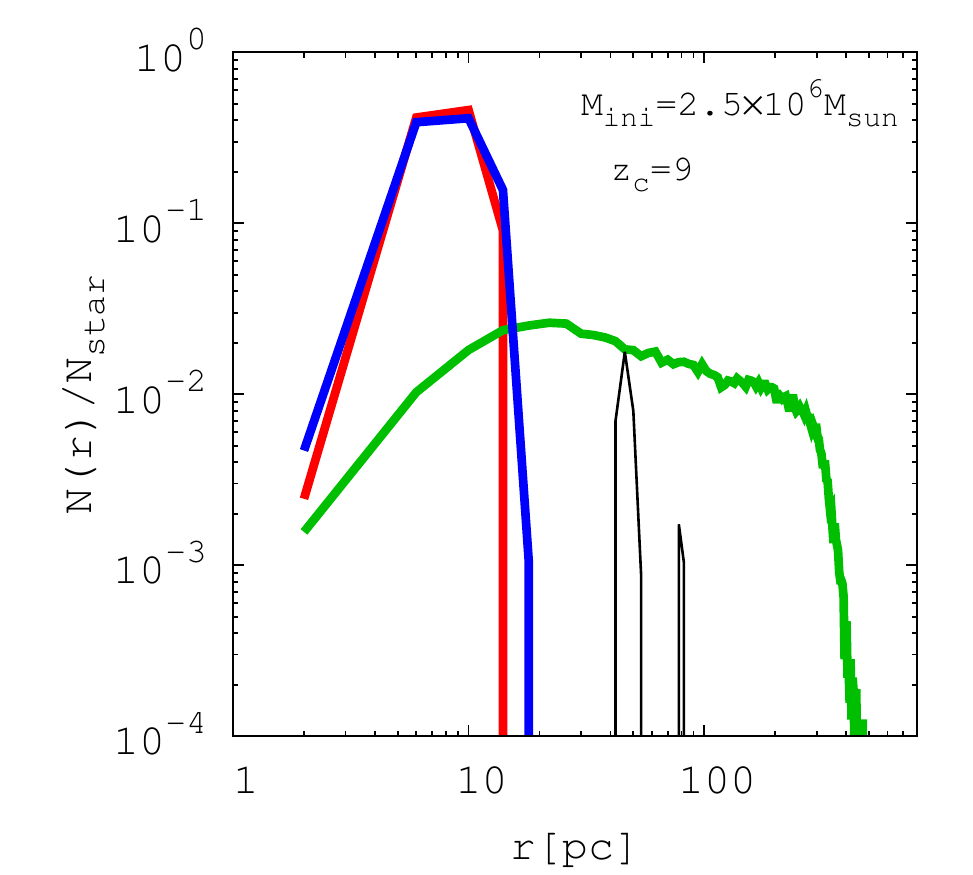}
		\end{center}
	\end{minipage}
	\begin{minipage}{0.33\hsize}
		\begin{center}
			\includegraphics[width=60mm,clip]{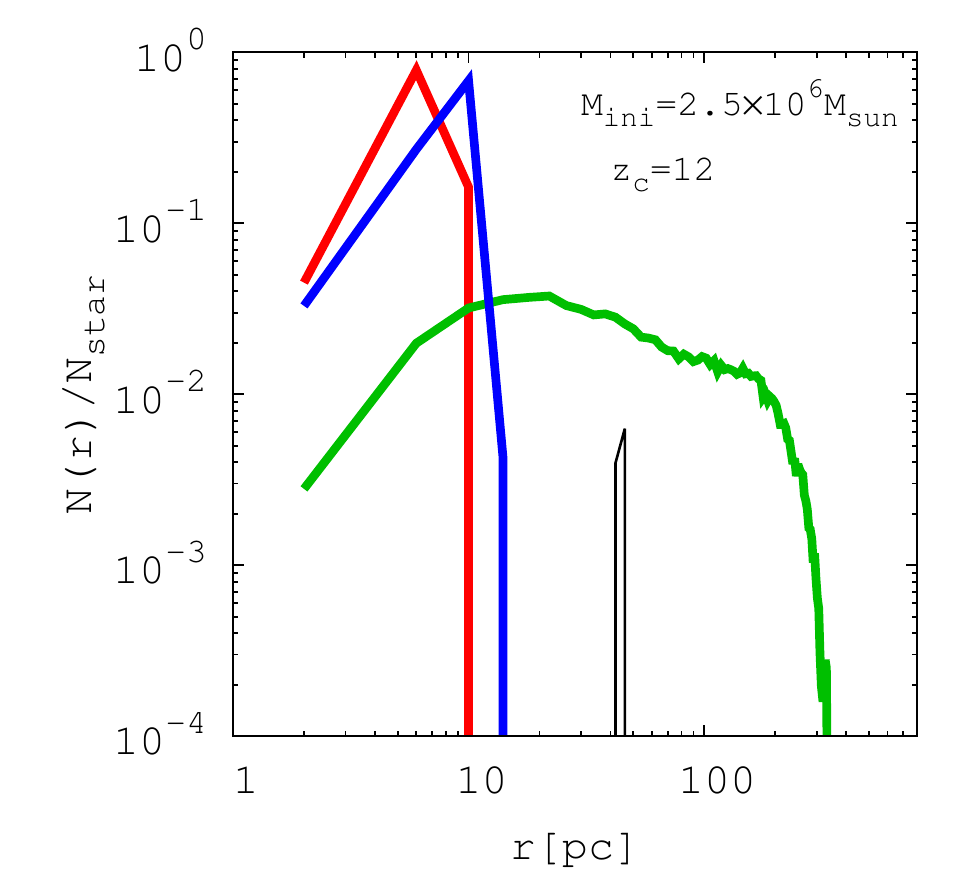}
		\end{center}
	\end{minipage}

	\begin{minipage}{0.33\hsize}
		\begin{center}
			\includegraphics[width=60mm,clip]{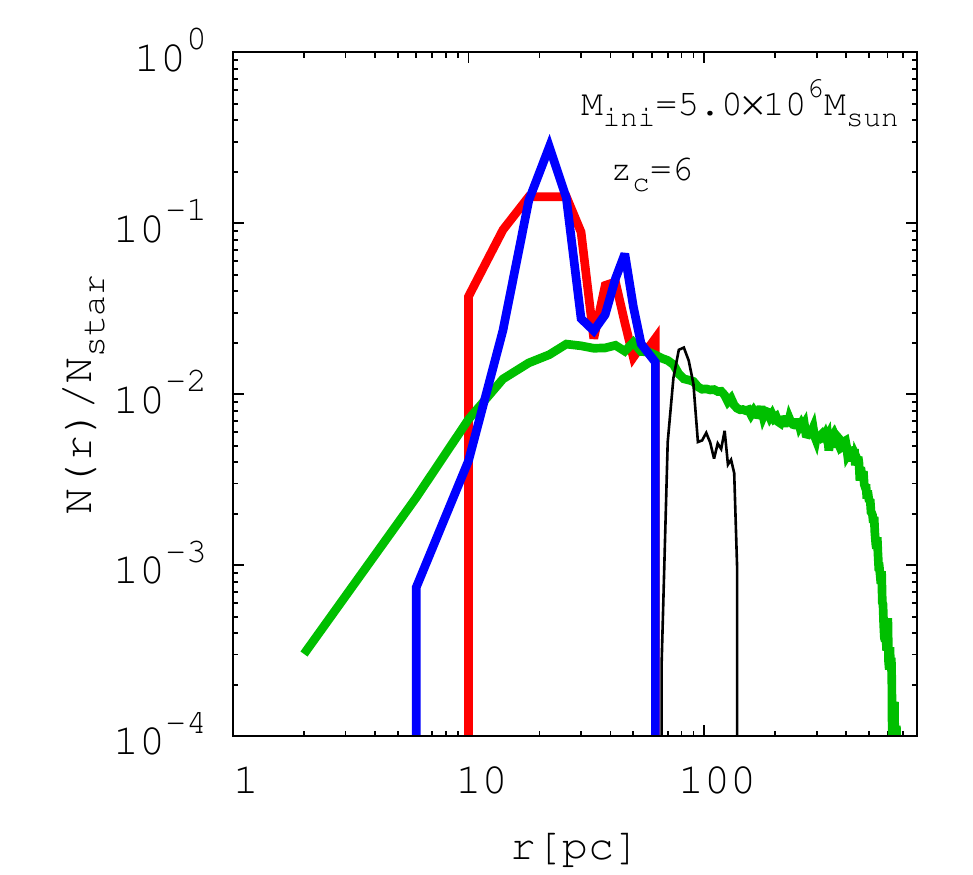}
		\end{center}
	\end{minipage}
	\begin{minipage}{0.33\hsize}
		\begin{center}
			\includegraphics[width=60mm,clip]{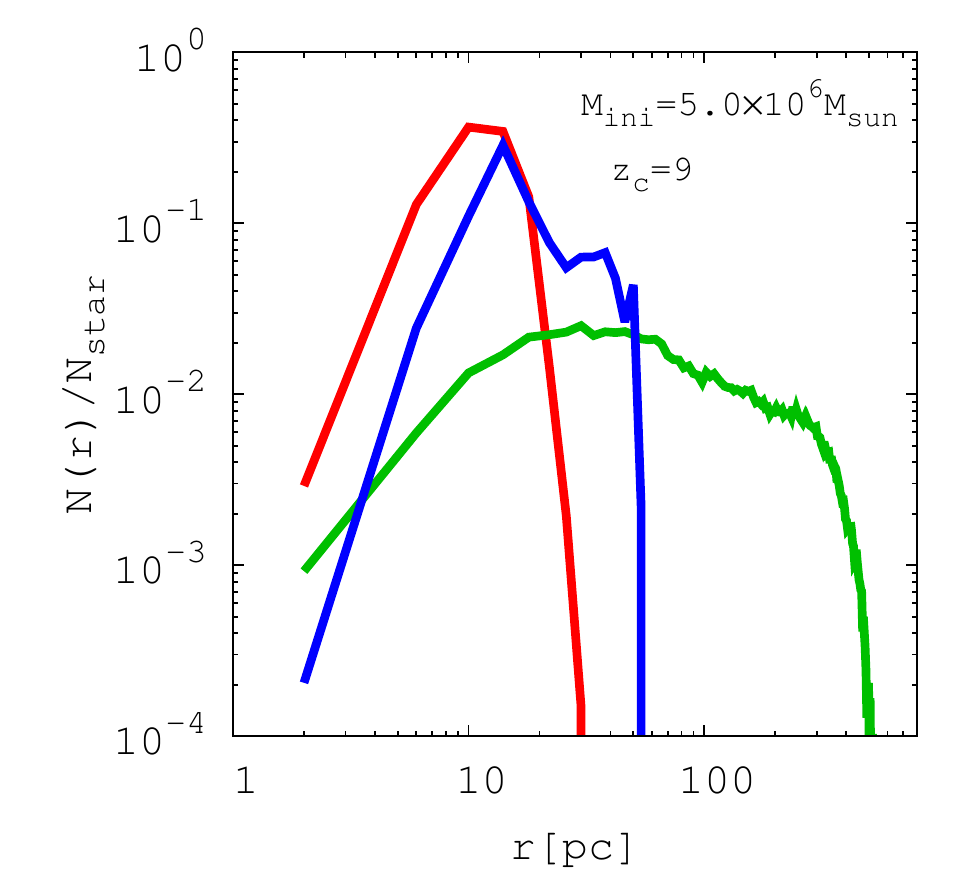}
		\end{center}
	\end{minipage}	
	\begin{minipage}{0.33\hsize}
		\begin{center}
			\includegraphics[width=60mm,clip]{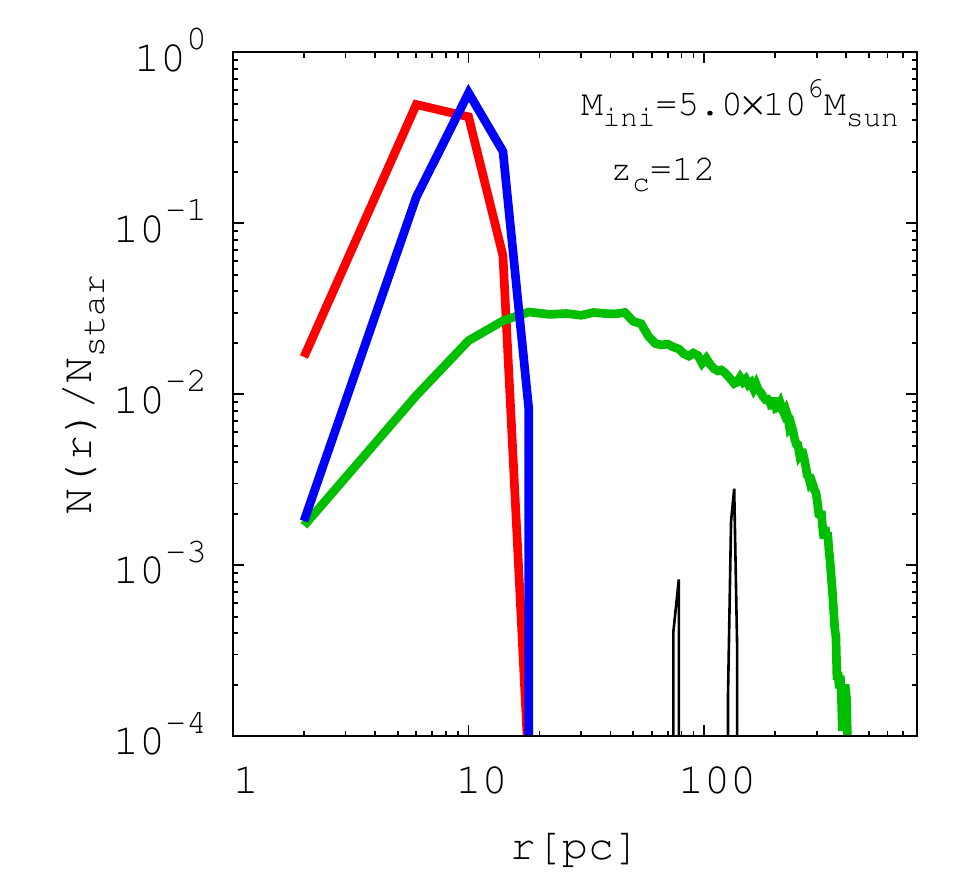}
		\end{center}
	\end{minipage}

		\caption{The distributions of the positions where SPH particles are converted to star particles. 
			The horizontal axis denotes the distance from the center of mass in units of pc, and
			the vertical axis does the number of formed stars normalized by the total number of stars.
			The upper three panels are the results for the cloud mass $M_{\rm ini} = 2.5\times 10^6\solmass$, 
			and the lower panels those for $M_{\rm ini} = 5\times 10^6\solmass$.
			From left to right, the panels show the models of $z_{\rm c}=6$, 9, and 12, respectively. 
			Each panel represents the results of ``supersonic infall'' 
under a strong one-sided background with $\dot{N}_{\rm ion}/\dot{N}_{\rm crit} = 10$ (red lines)
and a strong isotropic UV background with $\dot{N}_{\rm ion}/\dot{N}_{\rm crit} = 10$ (blue lines),
and the results of ``prompt star formation'' under a weak one-sided UV background 
with $\dot{N}_{\rm ion}/\dot{N}_{\rm crit} = 0.1$ (green lines). 
			Thin black lines in each panel show the distributions for star particles formed before UV irradiation. 
			} 
	\label{fig:position-distribution}
	\end{center}
\end{figure*}

\subsubsection{Star Formation History}\label{section:StarFormation}
As described in \S~\ref{SS}, the ionization and thermal properties in 3D-RHD calculations
are quite different from  those in 1D-RHD calculations. 
In particular, the difference of self-shielding between the one-sided and isotropic background radiation is noticeable. 
Furthermore, even though background radiation is isotropic, the $\HH$ distributions becomes more complicated 
owing to the inhomogeneous density fields in the cloud (e.g., the central column of Fig. \ref{fig:color-5M6-zcoll9-iso}). 
Therefore, it is expected that the star formation proceeds in a different fashion from that in the 1D-RHD calculations. 

To elucidate the three-dimensional effects, we firstly scrutinize the star formation history in each model.  
In Fig.~\ref{fig:position-distribution}, the positions where gas particles are converted to star particles are shown. 
In the runs of the ``supersonic infall'', we find that most stars form within several $10$pc 
from the center of the cloud after the UV irradiation (the red and blue lines in Fig.~\ref{fig:position-distribution}). 
This is due to the compactness of self-shielded regions formed by the strong UV background. 
We emphasize that the star-forming regions can be compact eventually even in the runs 
of anisotropic background radiation despite the extended shaded regions (e.g., Fig.~\ref{fig:color-5M6-zcoll9-1src}). 
This can be understood as follows; 
Although the shaded regions are immediately formed after the UV irradiation 
(e.g., low temperature regions in Fig. \ref{fig:1D-profile-5M6-zcoll9}), 
the photo-heated gas surrounds the shaded regions as time goes on,
as shown by the temperature map in Fig. \ref{fig:color-5M6-zcoll9-1src}. 
The broad shaded regions composed of cold gas are compressed by the photo-heated gas. 
But, photo-dissociating UV photons suppress $\HH$ formation, and therefore
stars cannot form in the shaded neutral regions until they are shoved by the hot gas 
toward the central part of the cloud and eventually shielded from photo-dissociating radiation
(Fig. \ref{fig:1D-profile-5M6-zcoll9}). 
As a result, the star forming regions become compact, even though the shaded regions 
are originally extended. 

\begin{figure*}
	\begin{center}

	\begin{minipage}{0.33\hsize}
		\begin{center}
			\includegraphics[width=60mm,clip]{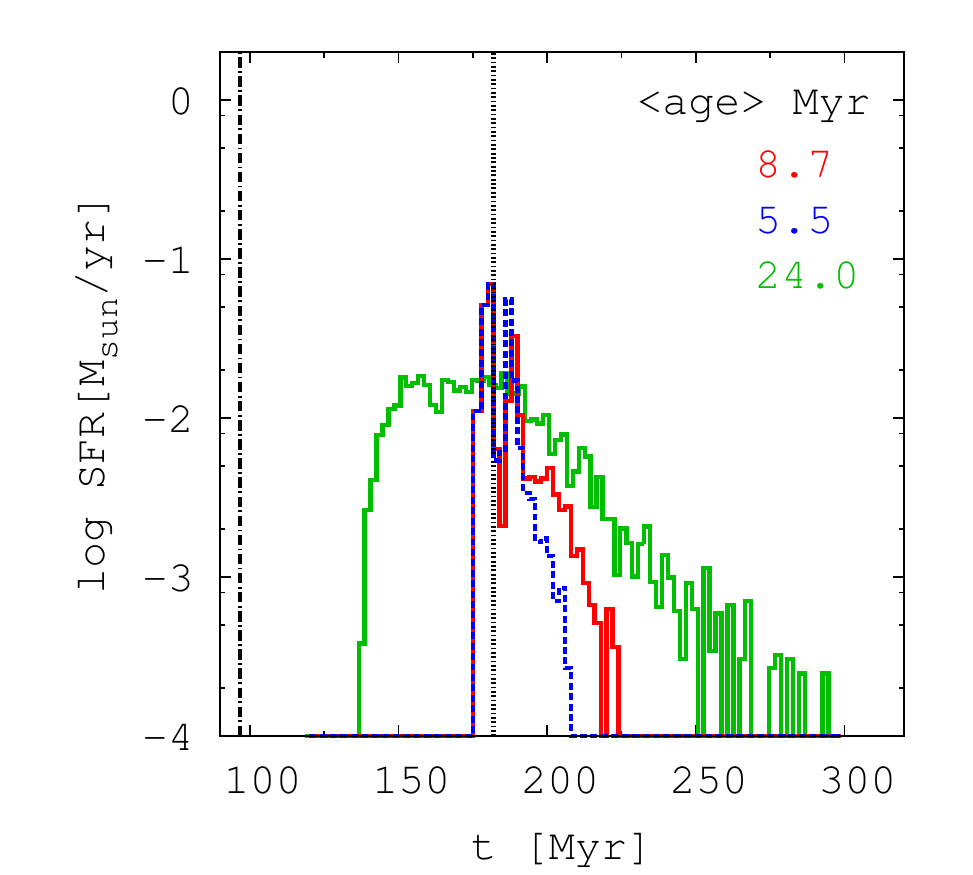}
		\end{center}
	\end{minipage}
	\begin{minipage}{0.33\hsize}
		\begin{center}
			\includegraphics[width=60mm,clip]{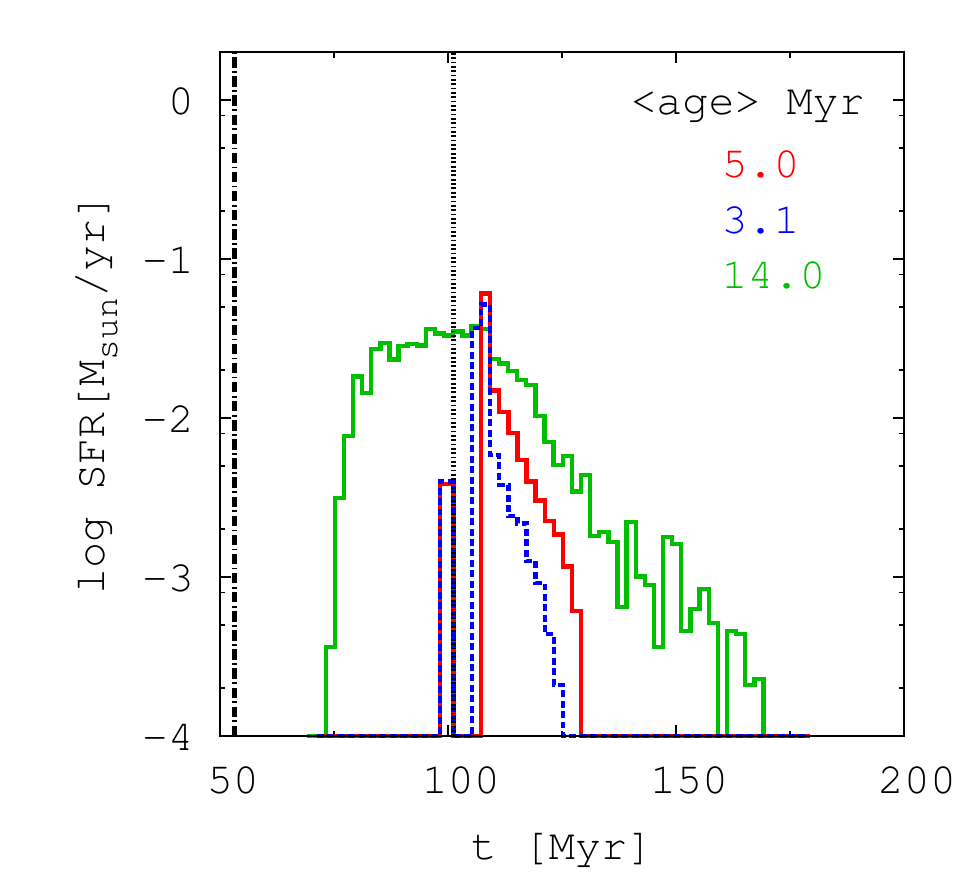}
		\end{center}
	\end{minipage}
	\begin{minipage}{0.33\hsize}
		\begin{center}
			\includegraphics[width=60mm,clip]{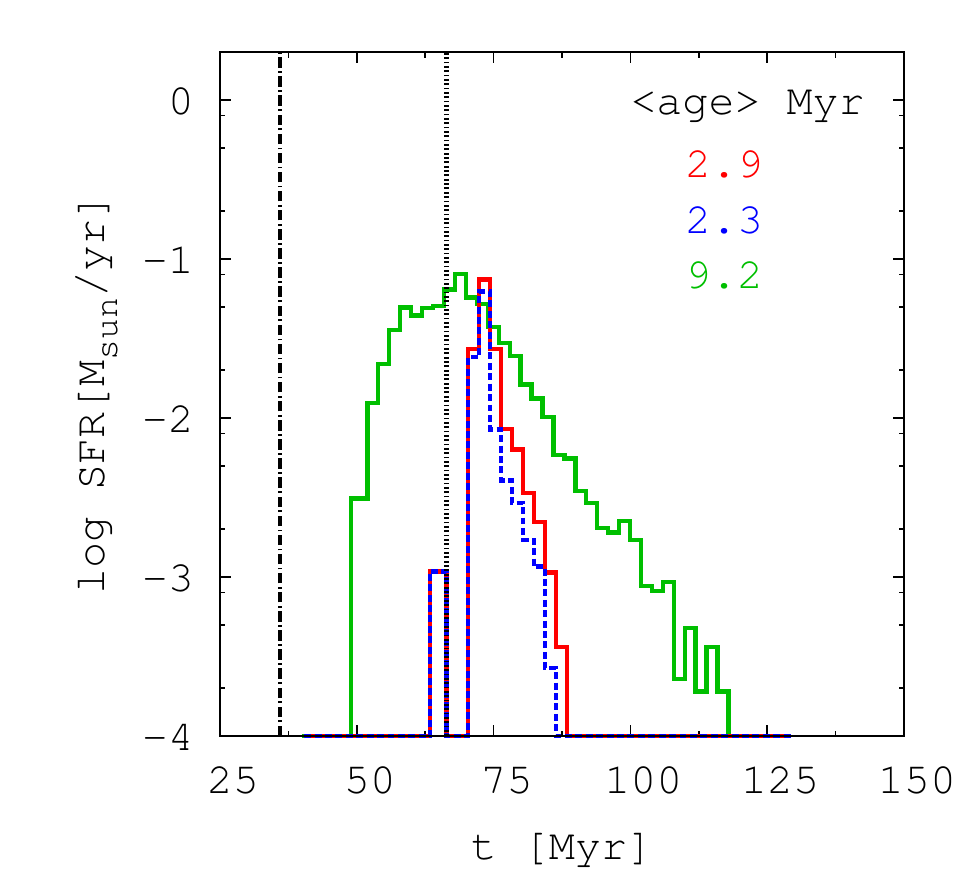}
		\end{center}
	\end{minipage}
		
	\begin{minipage}{0.33\hsize}
		\begin{center}
			\includegraphics[width=60mm,clip]{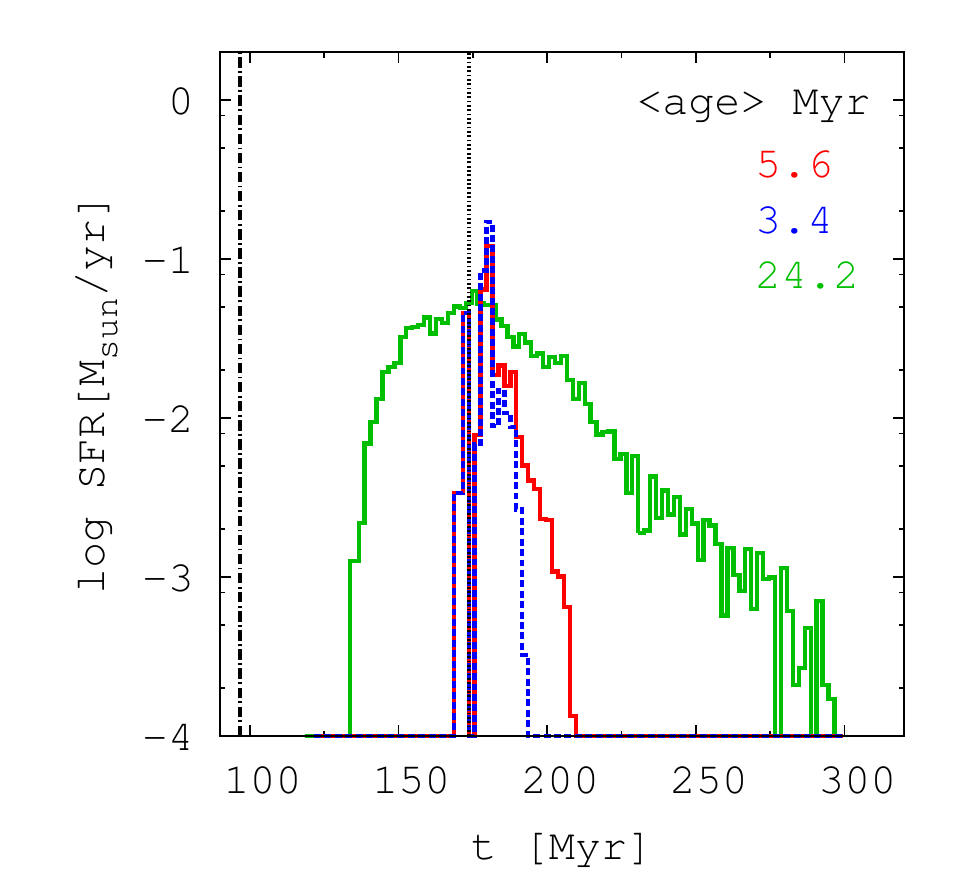}
		\end{center}
	\end{minipage}	
	\begin{minipage}{0.33\hsize}
		\begin{center}
			\includegraphics[width=60mm,clip]{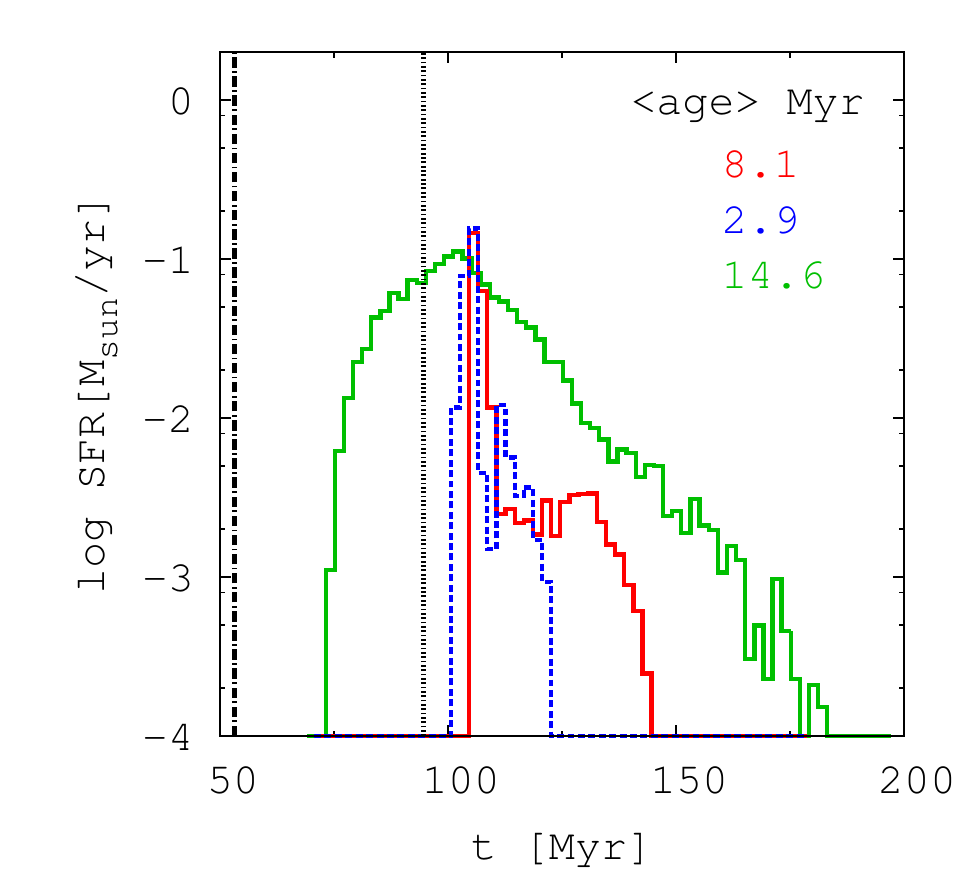}
		\end{center}
	\end{minipage}	
	\begin{minipage}{0.33\hsize}
		\begin{center}
			\includegraphics[width=60mm,clip]{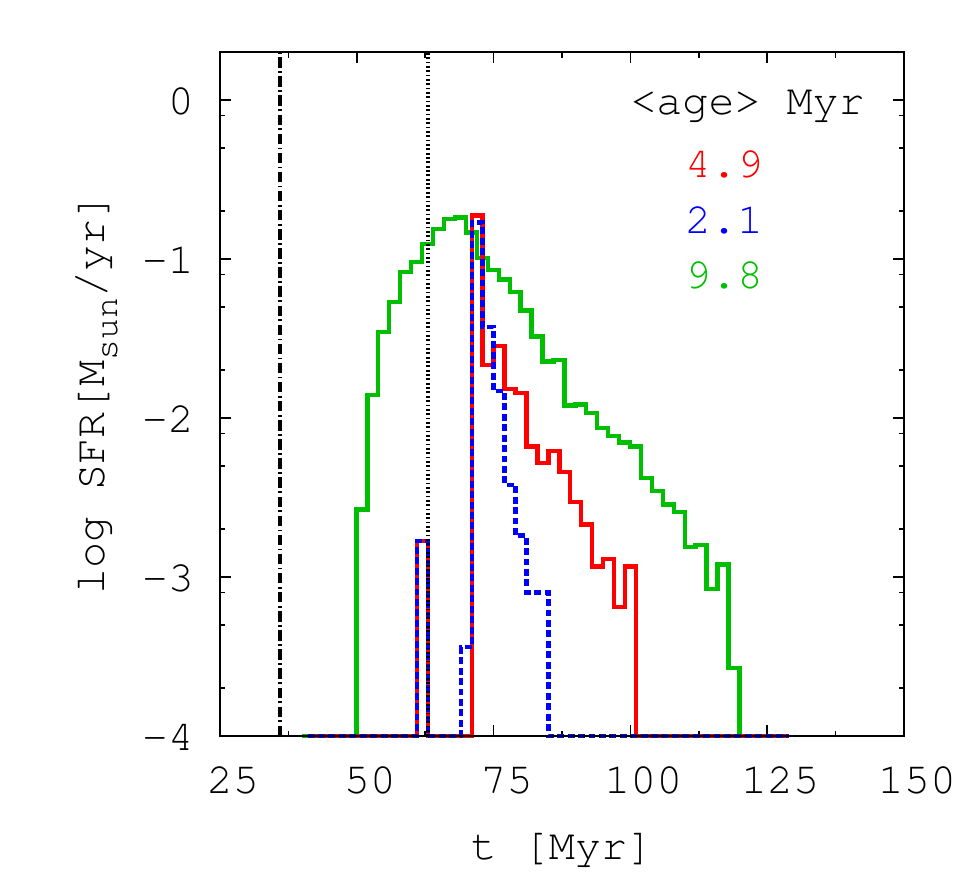}
		\end{center}
	\end{minipage}

		\caption{
The star formation rate in units of $\solmass$/yr as a function of time (Myr) for
each model presented in Fig.~\ref{fig:position-distribution}. 
			The red solid lines represent the runs of ``supersonic infall '' under an one-sided UV background,  
			while the blue dashed line represent those under an isotropic UV background. 
			The green solid lines denote the runs of ``prompt star formation''. 
			The vertical dotted lines in each panel indicate the epoch of the UV irradiation for the ``supersonic infall '',
			while the vertical dot-dashed lines indicate the epoch of the UV irradiation 
			for the ``prompt star formation''.  
            Also, in each panel, the age dispersions of all stars for each model are inserted.}
	\label{fig:SFR}
	\end{center}
\end{figure*}

However, the star formation history can be delayed by the anisotropy of UV background.  
We show the star formation rate as a function of time in Fig.~\ref{fig:SFR}. 
The typical duration of star formation in the ``supersonic infall'' runs is less than 10~Myr. 
Needless to say, such a short duration seems favorable to explain the single stellar population in GCs. 
Comparing the durations in the one-sided and isotropic UV background, 
we can see the typical duration in the one-sided UV background is slightly longer. 
Such delay is mainly caused by the star formation originating in the shaded regions. 
As already shown, the gas in the shaded regions is never photo-evaporated,
but is pushed inward by photo-ionized gas. 
The relatively slow infall of the shaded regions results in lengthening the duration of the star formation. 
 
The star formation in the ``prompt star formation'' model proceeds in a completely different fashion. 
In this case, the cloud is self-shielded promptly and the star formation begins 
shortly after the gravitational collapse. 
As shown by the green solid lines in Fig.~\ref{fig:position-distribution}, the star formation sites 
range from $\sim1$~pc up to several $100$~pc. 
The broad distributions in the ``prompt star formation'' originate in not only the extension of the self-shielded regions 
but also the positive feedback through $\HH$ formation around an ionization front, as mentioned in the previous section. 
Actually, the distributions of stars in the ``prompt star formation'' are more extended than 
the distributions of stars formed prior to the UV irradiation, as indicated by the black lines in Fig.~\ref{fig:position-distribution}. 
Also, the star formation history is quite different from that in the ``supersonic infall''. 
As shown in Fig. \ref{fig:SFR}, the star formation in the ``prompt star formation'' model continues over $\sim 100$ Myr. 
The long-term star formation is attributed to abundant material of self-shielded regions. 
Besides, stars can be induced by the positive feedback by ionizing photons in an earlier phase than a UV-free case.

\begin{figure*}
	\begin{center}	
		\begin{minipage}{0.33\hsize}
			\begin{center}
				\includegraphics[width=60mm,clip]{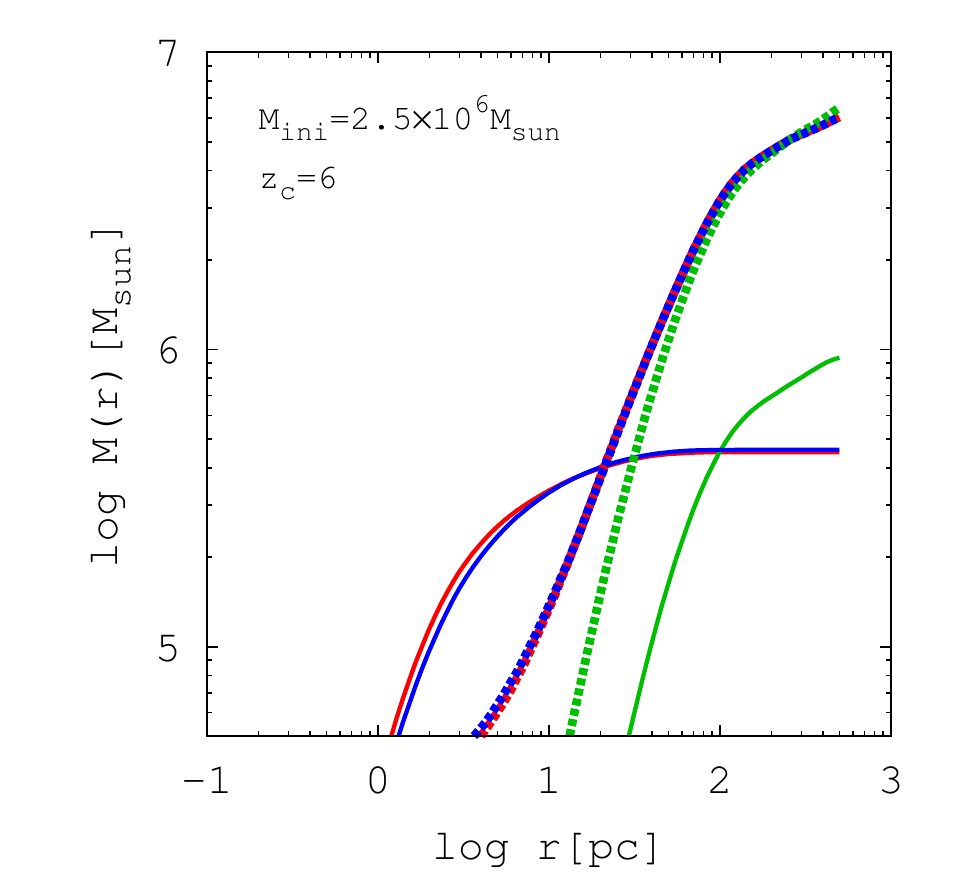}
			\end{center}
		\end{minipage}
		\begin{minipage}{0.33\hsize}
			\begin{center}
				\includegraphics[width=60mm,clip]{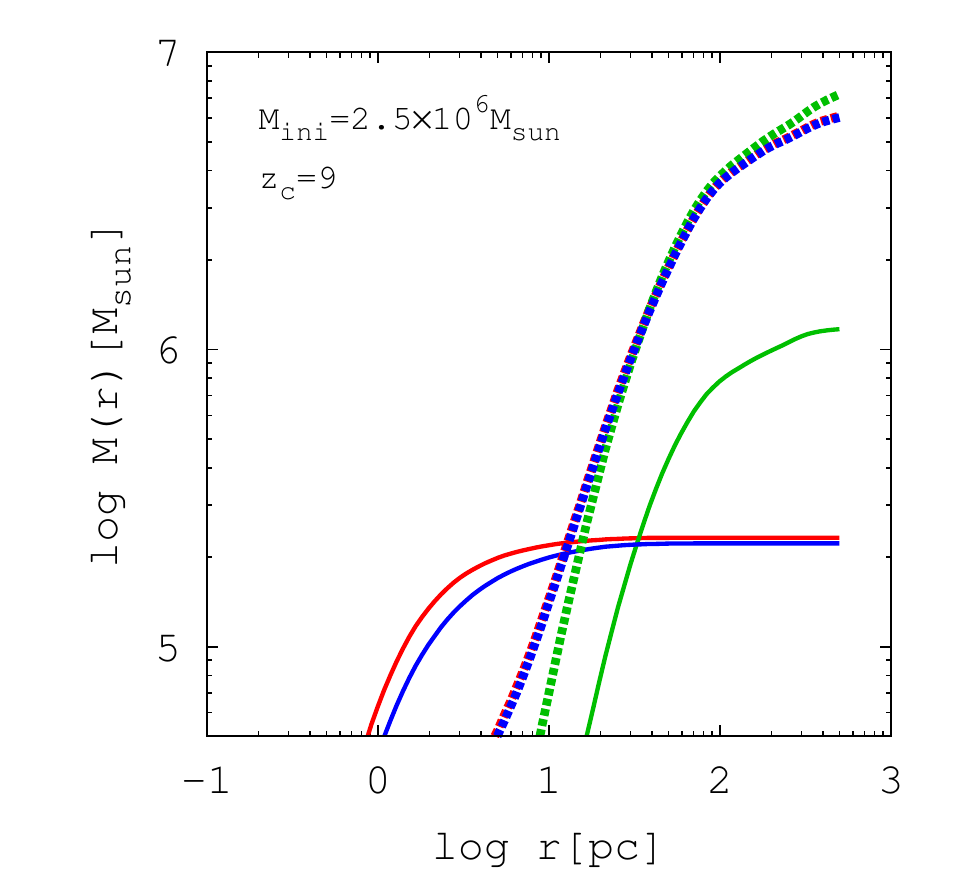}
			\end{center}
		\end{minipage}
		\begin{minipage}{0.33\hsize}
			\begin{center}
				\includegraphics[width=60mm,clip]{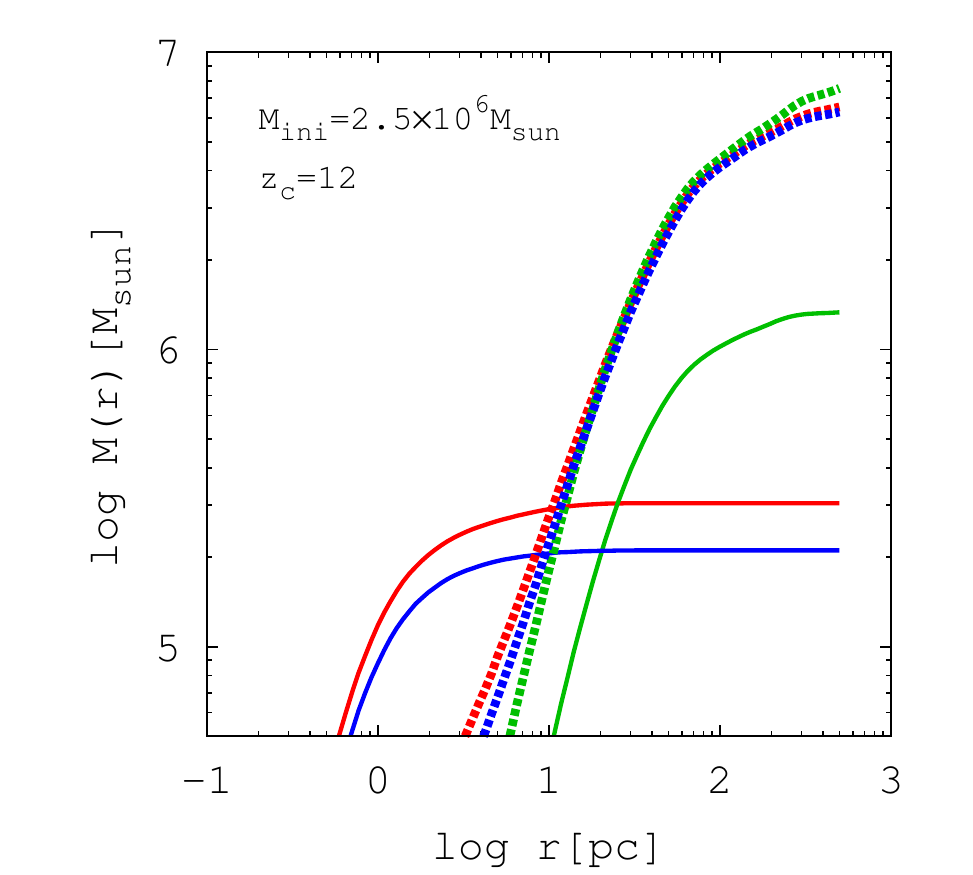}
			\end{center}
		\end{minipage}
		
		\begin{minipage}{0.33\hsize}
			\begin{center}
				\includegraphics[width=60mm,clip]{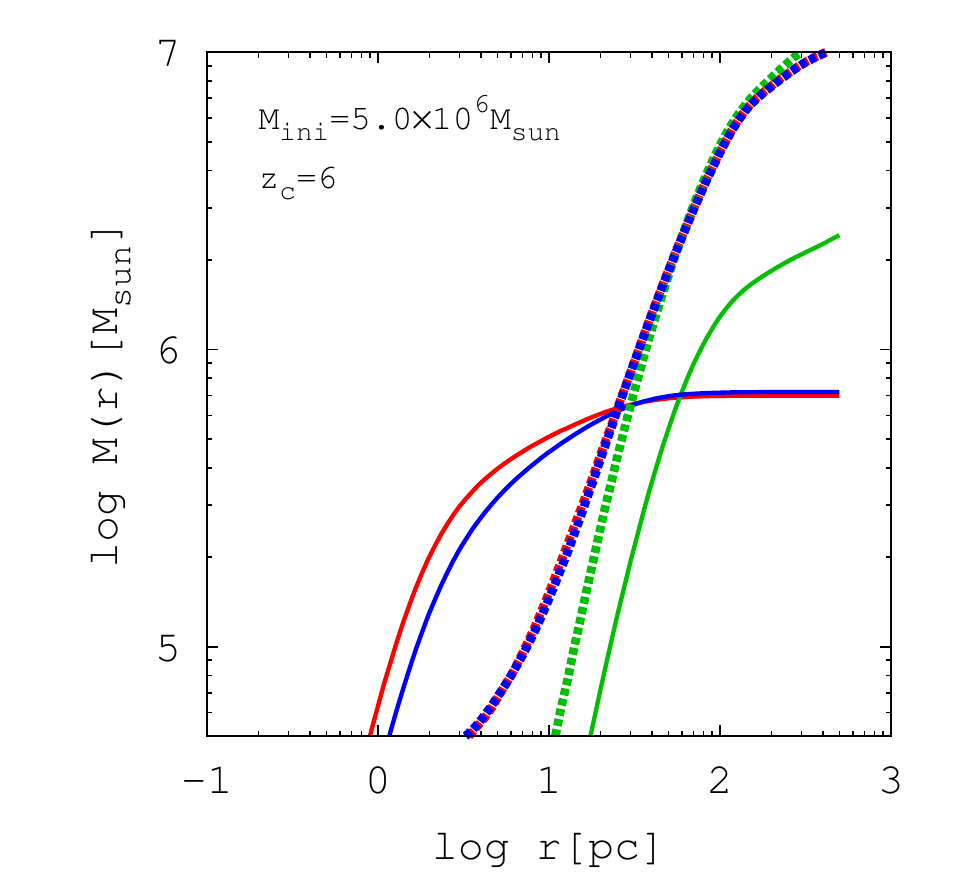}
			\end{center}
		\end{minipage}	
		\begin{minipage}{0.33\hsize}
			\begin{center}
				\includegraphics[width=60mm,clip]{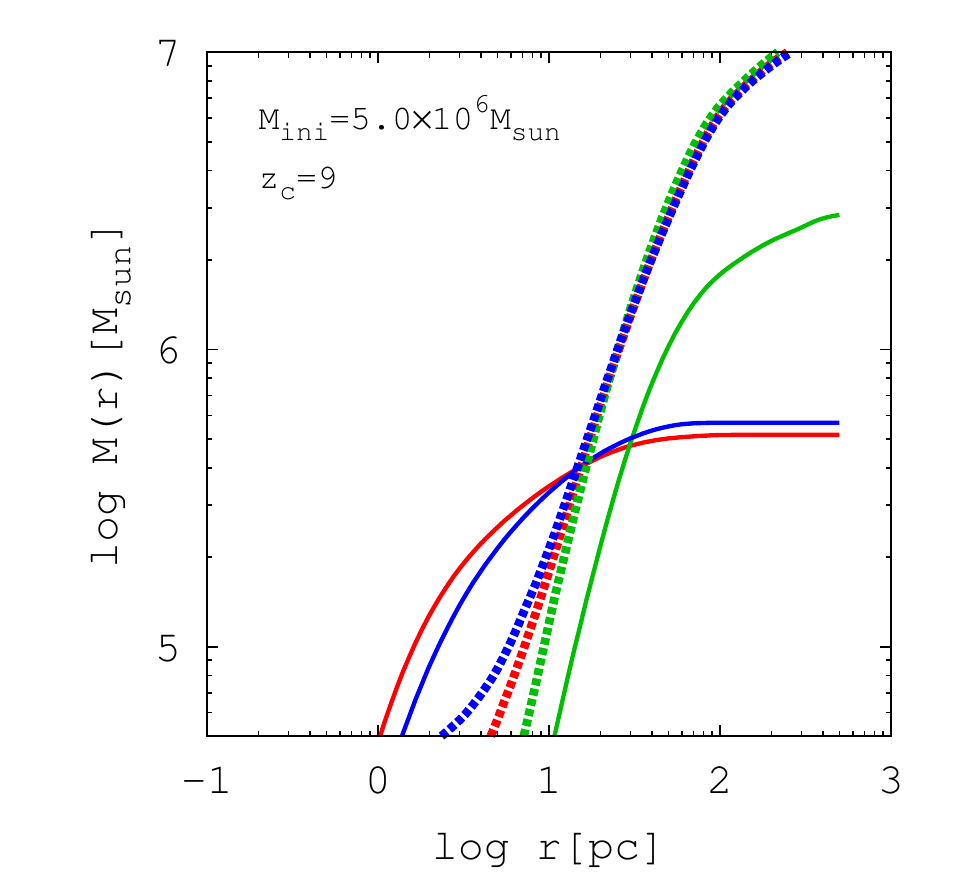}
			\end{center}
		\end{minipage}	
		\begin{minipage}{0.33\hsize}
			\begin{center}
				\includegraphics[width=60mm,clip]{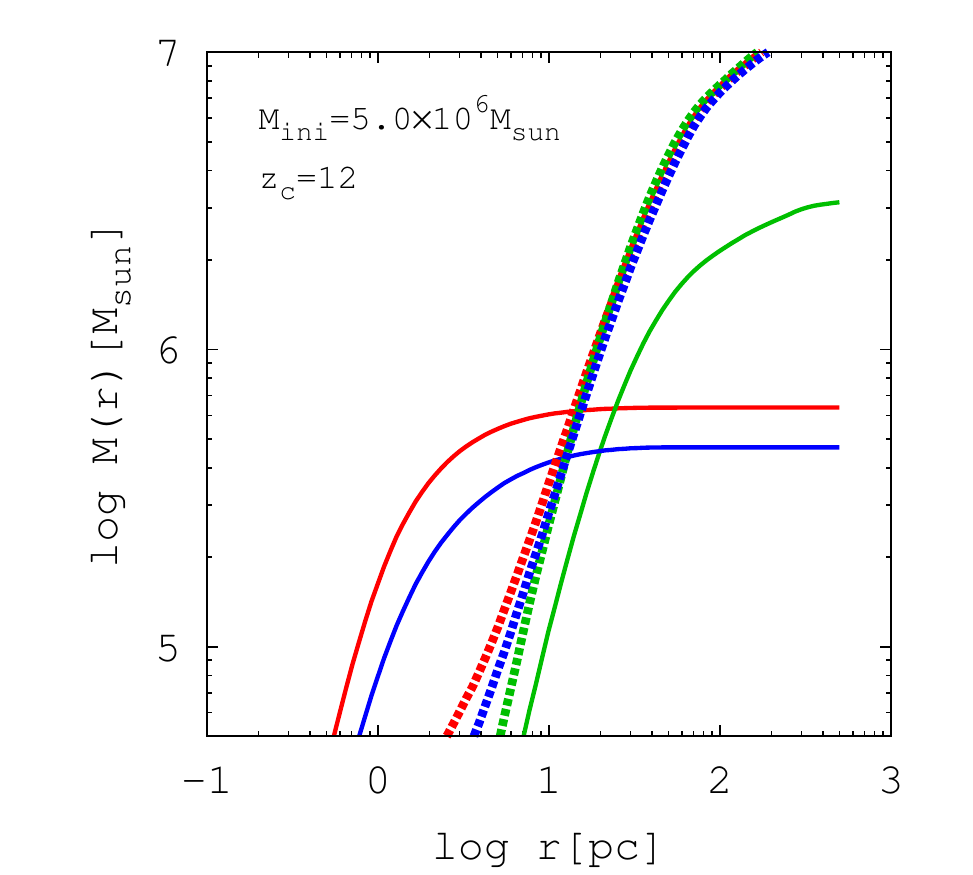}
			\end{center}
		\end{minipage}
		
		\caption{Mass distributions of stellar and dark matter components as a function of radii
for each model presented in Fig.~\ref{fig:position-distribution}.  
			The horizontal axis is the distance $r$ from the stellar density peak in units of pc, while 
			the vertical axis is the cumulative mass contained within $r$. 
			The solid and dotted lines respectively denote the stellar and dark matter components. 
			The red and blue lines respectively correspond to the one-sided and isotropic UV background
			in the ``supersonic infall'' model. 
			The green lines denote the ``prompt star formation'' model.  
}
	\label{fig:cumulative_mass}
	\end{center}
\end{figure*}

\subsection{Stellar Dynamics}
Here, we pursue the stellar dynamics until the simulated star clusters 
accomplish the quasi-steady state.
In this section, we describe the resultant features of the simulated star clusters. 

\subsubsection{Properties of Simulated Star Clusters}
Fig. \ref{fig:cumulative_mass} shows the cumulative mass profiles of the simulated star clusters. 
Obviously, the star clusters formed via ``supersonic infall'' become 
stellar-dominated in the main body of $\sim$ 10 pc (red and blue line). 
On the other hand, the star clusters formed through ``prompt star formation'' are dominated by dark matter (green line). 
The difference of the profiles reflects the way of contraction and star formation processes. 
In the ``prompt star formation'', stars form at an early phase of the cloud contraction without strong kinetic energy dissipation. 
Consequently, diffuse star clusters tend to form. 
In contrast, in the ``supersonic infall'', the thermal pressure enhanced by strong UV radiation 
dissipates the kinetic energy of contraction before stars are formed there. 
As a result, the formed star clusters tend to be compact and stellar-dominated. 
We here emphasize again that the properties of such compact star clusters are 
hardly affected by the anisotropy of radiation. 

To quantify the structure of the star clusters, we fit the stellar density profiles by the Plummer model,
which is known to be a good model for representing GC density profiles. 
The Plummer density profile is given by 
\begin{equation}
	\label{eq:Plummer}
		\rho(r) = \frac{3M_\ast}{4\pi b^3}\left(1+\frac{r^2}{b^2}\right)^{-5/2}, 
\end{equation}
where $M_\ast$ denotes the total stellar mass of a cluster and $b$ denotes the Plummer scale length
that corresponds to the core radius of the cluster. 
The core radii fitted for all of the simulated clusters are listed in Table \ref{table:core_radius}. 
As seen in Table \ref{table:core_radius}, the core radii of the star clusters formed via ``supersonic infall'' 
are well concordant with the core radii of $\lesssim 1$ pc observed in GCs \citep[e.g.,][]{Kormendy85}. 
Also, the difference between the one-sided and isotropic UV background is small.
On the other hand, the typical core sizes of the clusters formed via ``prompt star formation'' 
are much larger than $10$ pc. \footnote{
We should note that the core radii change as the clusters dynamically evolve via the two-body relaxation.
To make more precise comparison between the simulated clusters and observed GCs, 
the effects of two-body relaxation should be carefully incorporated
in the collisional $N$-body simulations, which will be explored elsewhere.}

\begin{table*}
\begin{center}
\caption{The core radii of simulated star clusters}
\label{table:core_radius}
\begin{threeparttable}
\begin{tabular}{ccc|ccc|ccc|ccc}

\hline  \hline
\multicolumn{3}{c|}{}&
\multicolumn{3}{c|}{supersonic/one-sided} & 
\multicolumn{3}{c|}{supersonic/isotropic} & 
\multicolumn{3}{c}{prompt star formation} 
\\
\hline
$z_{\rm c}$ & 
$M_{\rm ini} $ & 
$t_{\rm rise}$ &
$b$  &
$M_{\rm core} $ \tnote{1} &
$M_\ast/M_{\rm ini}$ \tnote{2} &
$b$  &
$M_{\rm core} $ &
$M_\ast/M_{\rm ini} $&
$b$&
$M_{\rm core} $ & 
$M_\ast/M_{\rm ini} $\\
&
[$10^6\solmass$] & [yr] &
[pc] &
[$10^5\solmass$] & &
[pc] &
[$10^5\solmass$] & &
[pc] &
[$10^5\solmass$] 
\\
\hline
    6 & 2.5	& instant 	& 1.7 & $0.93$  & 0.18		& 1.8 & $0.85$ & 0.18		& 55.9 & $2.0$ & 0.40 \\
    6 & 5.0 	& instant 	& 1.6 & $1.5$  & 0.14		& 2.0 & $1.3$ & 0.14			& 46.7 & $4.8$ & 0.53\\
    9 & 2.5  & instant	& 0.94 & $0.57$  & 0.093		& 1.0 & $0.47$ & 0.089 		& 33.7 & $2.3$ & 0.49\\
    9 & 5.0 &  instant	& 1.4 & $0.85$  & 0.10 		& 1.8 & $0.77$ & 0.11		& 30.4 & $5.0$ & 0.60 \\
    9 & 5.0 & $10^6$	& 1.5 & 0.85  & 0.12 		& - & - & -		& - & - & - \\
    9 & 5.0 & $10^7$	& 2.4 & 1.6  & 0.23		& - & - & -		& - & - & - \\
    9 & 5.0 & $10^8$	& 6.7 & 16.8  & 0.42 		& - & - & -		& - & - & - \\
    12 & 2.5 & instant	& 0.78 & $0.83$ & 0.12		& 0.76 & $0.60$ &  0.084		&22.2 & $2.5$  & 0.55\\
    12 & 5.0 & instant	& 1.0 & $1.7$  & 0.13		& 1.2 & $1.1$  & 0.093		& 21.0 & $5.0$ &  0.64\\
     9 & 1.0 & instant	& - & - & - 					& - & - & - 					& 32.6 & $0.58$ & 0.32 \\
     9 & 10.0 & instant	& 2.1 & $3.2$ & 0.20 		& - & - & - 					&  25.8 & $8.6$ & 0.69 \\
    12 & 1.0 & instant 	& - & - & - 					& - & - & - 					& 25.0 & $0.83$ & 0.41 \\
    12 & 10.0 & instant	&1.7 & $4.4$ & 0.30 			& - & - & - 					& - & -& - \\
\hline
\end{tabular}
	\begin{tablenotes}
		\item[1] Stellar mass within the scale length $b$ in units of $10^5\solmass$
		\item[2] The ratio of total stellar mass to the initial gas mass
	\end{tablenotes}
\end{threeparttable}
\end{center}
\end{table*}

\begin{figure}
	\begin{center}
		\includegraphics[width=100mm,clip]{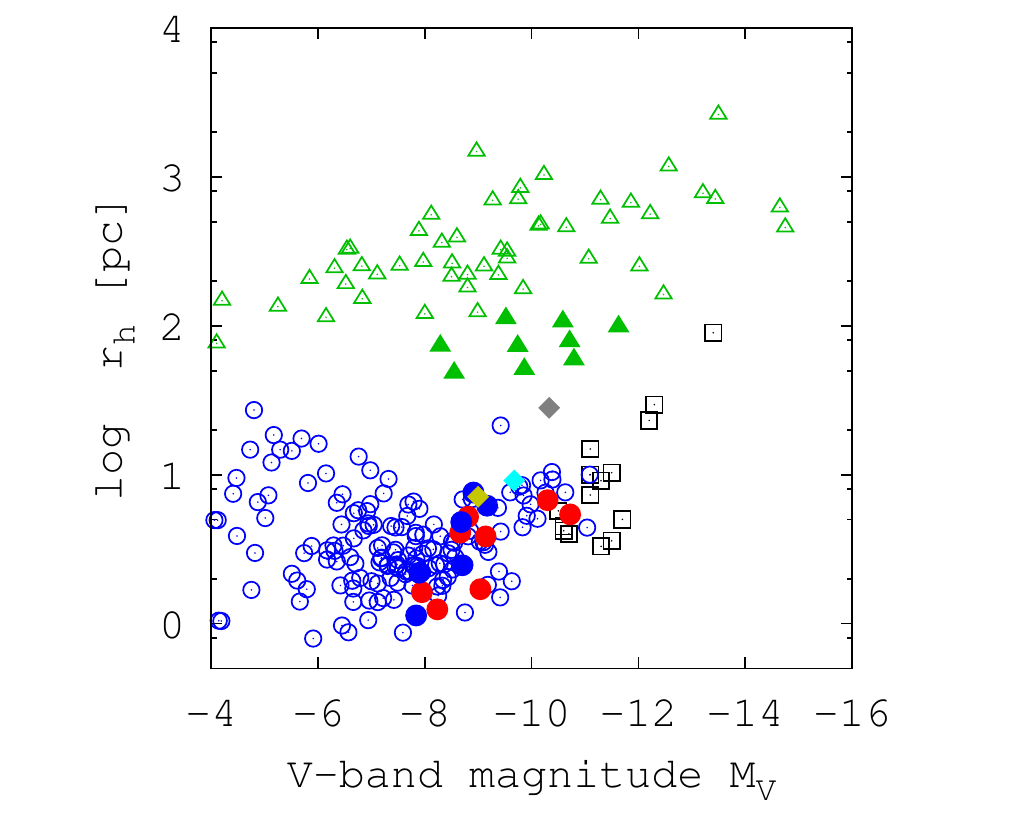}
		\caption{Half-mass radii $r_{\rm h}$ of star clusters as a function of absolute V-band magnitude $M_{\rm V}$. 
			The filled red and blue circles indicate the star clusters formed through ``supersonic infall'' in the one-sided and isotropic UV background, respectively. 
			The filled  diamonds represent the star clusters formed in the time-evolving one-sided UV background, and the colors of yellow, cyan, and gray correspond to the linear rising time of $t_{\rm rise} = 10$ Myr, 10 Myr, and 100 Myr, respectively. 
			The filled green triangles are the star clusters formed via ``prompt star formation''. 
			The open circles, triangles and squares denote the observed GCs, dSphs and UCDs, respectively. 
			The observational data for GCs are taken from the MW GCs catalog of \citep{Harris1996} and NGC 5218 GCs \citep{Martini&Ho04}. 
			Those for dSphs are taken from \citet{McConnachie12}, and for UCDs are taken from \citet{Drinkwater} and \citet{Mieske+08}. 
			The thick solid line indicates a relationship of $\sigma \propto L^{1/3}$. }
			\label{fig:Rh-Mv}
	\end{center}
\end{figure}
\begin{figure}
	\begin{center}
		\includegraphics[width=100mm,clip]{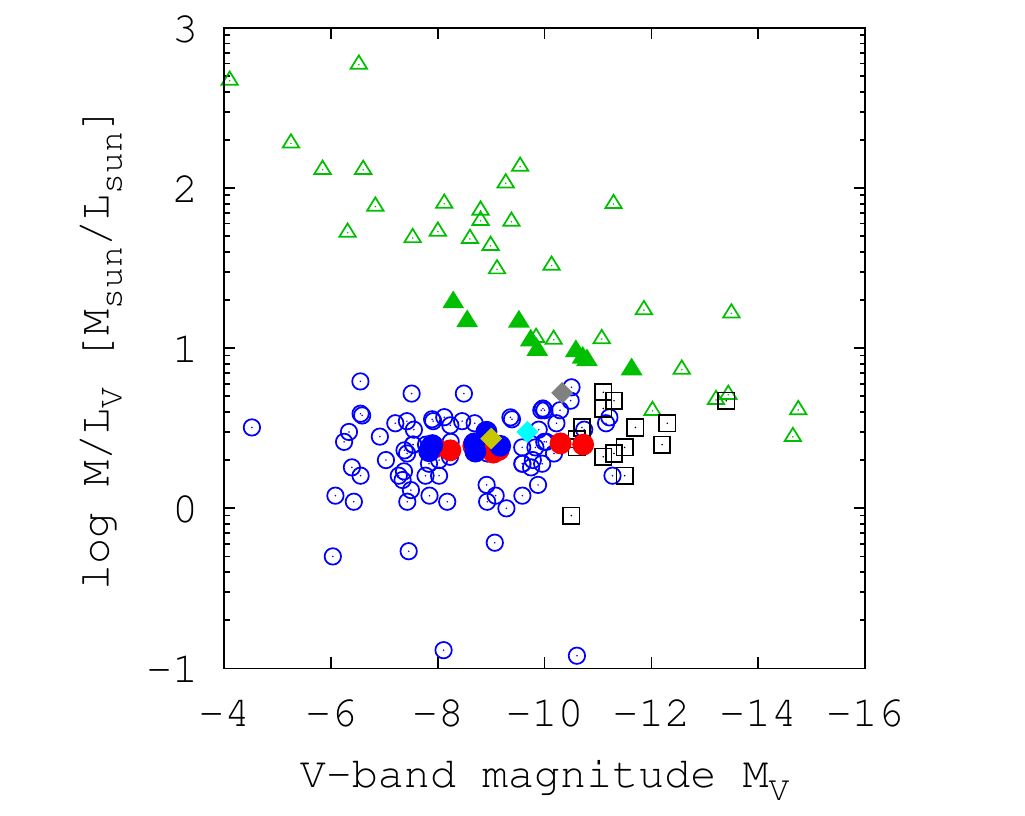}
		\caption{Mass-to-light ratios $M/L_{\rm V}$ as a function of absolute V-band magnitude $M_{\rm V}$. 
			The meanings of the symbols are the same as Figure \ref{fig:Rh-Mv}. 
			The observational data for GCs are taken from MW GCs, 
			LMC GCs,
			SMC GCs, 
			Fornax GCs, 
			NGC 5218 GCs, 
			and M31 GCs, compiled by HUK09.  
			The observational data for dSphs are taken from \citet{McConnachie12}. 
			As for UCDs, the data are taken from \citet{Drinkwater} and \citet{Mieske+08}. }
			\label{fig:ML-Mv}
	\end{center}
\end{figure}
\subsubsection{Comparison to Observations}\label{sec:comparison}
In this section, we attempt to compare the simulated star clusters to observations.
For the comparison, the V-band magnitude $M_{\rm V}$ of the simulated clusters is derived by assuming the typical mass-to-light ratio for GCs as ${M_{\rm GC}}/{L_V}= 2$ \citep{Pryor&Meylan93}. 

Fig. \ref{fig:Rh-Mv} shows the resultant half-mass radii $r_{\rm h}$ for the stellar components of the simulated clusters as a function of $M_V$. Those of the observed GCs, dSphs, and UCDs are also plotted in the figure. 
We find that the star clusters formed through ``supersonic infall'' exhibit $r_{\rm h}\sim$ 1-10 pc, as observed GCs show. 
The compactness originates in their compact star-forming regions as well as strong kinetic energy dissipation. 
The result indicates that strong UV radiation is one of the keys to reproduce small half-mass radii of GCs. 
In contrast, the star clusters formed via ``prompt star formation'' are never distributed 
around the observed GCs on the $r_{\rm h}-M_{\rm V}$ plane. 
The half-mass radii of the ``prompt star formation'' star clusters are higher by an order of magnitude  
than those of the ``supersonic infall'' star clusters at any $M_{\rm V}$.

In Fig. \ref{fig:ML-Mv}, we compare the mass-to-light ratios $M_{\rm dyn}/L_{\rm V}$ of the simulated clusters with those of GCs. 
We define the dynamical mass $M_{\rm dyn}$ by the total mass within the half-mass radius 
$M_\ast(r<r_{\rm h}) + M_{\rm DM}(r<r_{\rm h})$. 
As shown in \ref{fig:ML-Mv}, the mass-to-light ratios of the star clusters formed via ``supersonic infall''
are consistent with those of GCs. 
 On the other hand, the mass-to-light ratios of the star clusters in the ``prompt star formation'' are typically $\sim$ 10,
which are considerably higher than those of GCs. 
In other words, the star clusters formed through ``prompt star formation'' inevitably become dark matter-dominant systems. 
They seem to belong to the class of dSphs rather than stellar-dominated systems such as GCs or UCDs.  

\begin{figure}
	\begin{center}
		\includegraphics[width=100mm,clip]{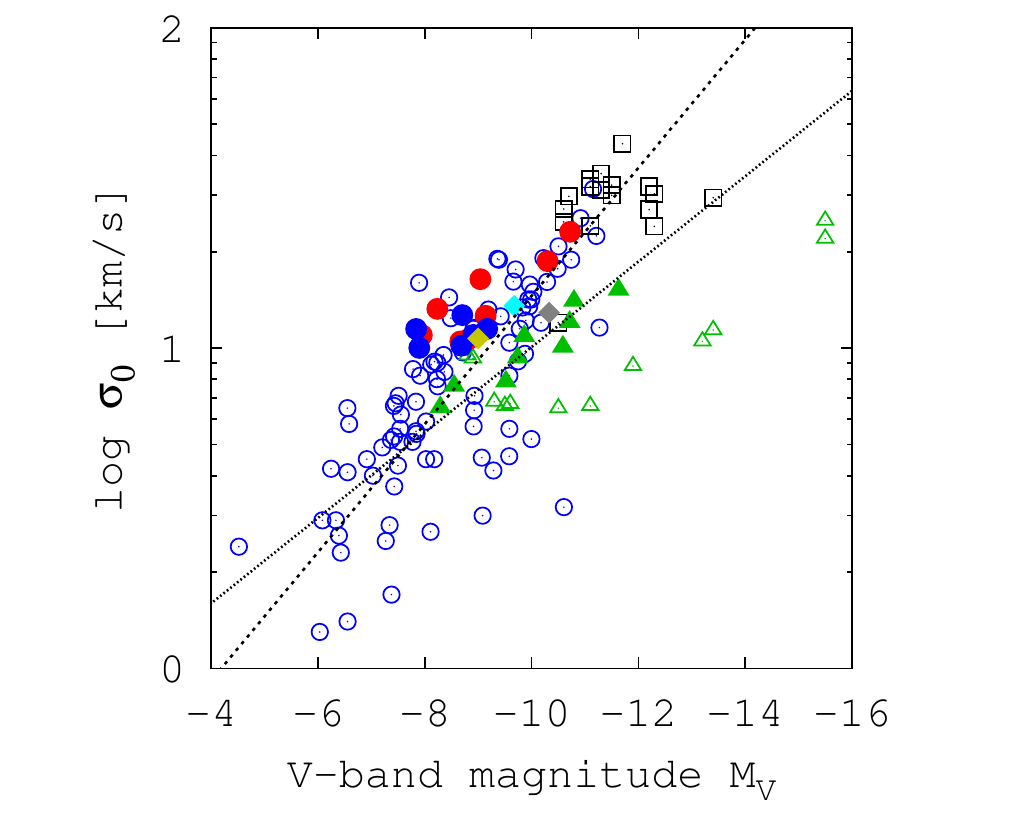}
		\caption{Central velocity dispersions $\sigma_0$ as a function of absolute V-band magnitude. 
			The meanings of the symbols are the same as Fig.~\ref{fig:Rh-Mv}. 
			The observational data for GCs and UCDs are taken from the literatures shown in Fig.~\ref{fig:ML-Mv}. 
			The observational data for dSphs are taken from \citet{Mateo98}. 
			The dotted line represents the best-fitted relation of $\sigma \propto L^{1/2}$ for GCs derived by \citet{Hasegan}. 
			The solid line indicates a relation of  $\sigma \propto L^{1/3}$. }
			\label{fig:sigma-Mv}
	\end{center}
\end{figure}

In Fig.~ \ref{fig:sigma-Mv}, we show the central velocity dispersions $\sigma_0$ 
of the simulated star clusters as a function of $M_V$ and compare them with observations. 
As seen in Fig.~ \ref{fig:sigma-Mv}, the star clusters in the ``supersonic infall'' 
result in higher velocity dispersions than those in the ``prompt star formation'' at given $M_{\rm V}$. 
As for the ``prompt star formation'', some clusters show high velocity dispersions of $\sim 10$ km/s, 
but they are mainly determined by dark matter potential rather than stars. 
According to the virial theorem, a velocity dispersion is roughly expressed 
by using the total mass $M$ and the half-mass radius $r_{\rm h}$ of a star cluster as
\begin{equation}
	\sigma_0 \sim \sqrt{ \frac{GM}{r_{\rm h}} }. 
\label{sigma}
\end{equation}
If $M \propto r_{\rm h}^3$ and $L\propto M$ are assumed, equation (\ref{sigma}) gives the relation of $\sigma_0 \propto L^{1/3}$.
Observed GCs, however, show the relation like $\sigma_0 \propto L^{1/2}$ \citep{Hasegan}.
These two relations are shown in Fig.~ \ref{fig:sigma-Mv}.
The velocity dispersions of the star clusters formed through ``supersonic infall'' do not satisfy $\sigma_0 \propto L^{1/3}$,
but exhibit higher values as those of GCs. 
However, it is not clear whether the simulated clusters obey the relation of $\sigma \propto L^{1/2}$,
since the number of the sample simulated clusters is not enough. 
The relation of $\sigma_0 \propto L^{1/2}$ is satisfied, only if the size of the system is almost independent of the mass.  
The half-mass radii of the present cluster samples slightly depend on the stellar mass in Fig.~\ref{fig:Rh-Mv}. 
To argue $\sigma_0-M_{\rm V}$ relation more quantitatively, probably we should consider carefully
other processes such as tidal stripping by host galaxies and internal feedback, which are discussed below in \S~\ref{sec:dis}. 
These additional processes seem to be important to reproduce low-mass GCs as well that are not presented in this work. 
Although the simulations of such additional effects will be left for the future work,
the properties of the simulated star clusters on the diagrams in Figs.~\ref{fig:Rh-Mv}, \ref{fig:ML-Mv}, and \ref{fig:sigma-Mv} 
match those shown in 1D simulations by HUK09. 
We emphasize that the combination of strong UV background radiation and ``supersonic infall'' 
provides a potential mechanism for the formation of compact star clusters as observed GCs. 

\section{Discussion}\label{sec:dis}
\subsection{Formation Sites of Globular Clusters}
\begin{table}
\begin{center}
\caption{Ionizing photon flux required for supersonic infall star formation}
\label{table:UVbackground}
\begin{tabular}{cccc}
\hline \hline
$z_{\rm c}$ & 
$z_{\rm UV}$ &
$M_{\rm ini} $ & 
$F_{\rm ion} $  \\
& &
[$10^6M_\odot$] & 
[${\rm photons}\: {\rm cm^{-2}}\: {\rm s^{-1}}$] 
\\
\hline 
    6 &  6.8 & 2.5  & $7.3\times 10^8$ \\
    6 & 6.9 & 5.0 & $7.6\times 10^8$ \\
    9 & 10.3 & 2.5 & $1.8 \times 10^9$ \\
    9 & 10.5 & 5.0 & $4.5 \times 10^8$ \\
    9 & 10.5 & 10.0  & $1.4 \times 10^9$ \\
    12 & 13.8 & 2.5  & $3.3 \times 10^9$ \\
    12 & 14.0 & 5.0  & $2.2 \times 10^9$ \\
    12 & 14.0 & 10.0  & $4.0 \times 10^9$ \\
\hline
\end{tabular}
\end{center}
\end{table}

As we have seen in the previous sections, strong UV background radiation is 
one of the essential conditions to produce GC-like compact star clusters. 
In addition, we can recognize in Figs.~\ref{fig:Rh-Mv}-\ref{fig:sigma-Mv} that 
the timescale $t_{\rm rise}$ of UV intensity rise is a significant factor for the cluster formation.
If $t_{\rm rise} \lesssim$10 Myr, then the background UV intensity reaches the maximum value
before the cloud undergoes the extensive star formation. 
As a result, the evolution of gas clouds differs little from the case of constant UV background, 
resulting in the ``supersonic infall''. 
On the other hand, if the rise of the background UV is as slow as $t_{\rm rise} >$10 Myr,  
the ``prompt star formation'' proceeds instead of ``supersonic infall'', 
since the self-shielding is effective in an early phase of contraction. 
Consequently, the results deviate from ``supersonic infall'' , 
as shown by gray diamonds in Figs.~\ref{fig:Rh-Mv}-\ref{fig:sigma-Mv}. 
Therefore, the rise of UV radiation should be faster than the cloud contraction to form GCs. 
We argue the formation sites of GCs  from viewpoints of UV radiation intensity and its variation timescale. 

For the purpose, we firstly evaluate the photon number flux required. 
We define $F_{\rm ion}$ as 
\begin{equation}
	F_{\rm ion} \equiv \frac{\dot{N}_{\rm ion}}{\pi r_{\rm UV,in}^2}, 
\label{F_ion}
\end{equation}
where $r_{\rm UV,in}$ denotes the radius of a cloud at the irradiation epoch. 
Although this estimation is higher by a factor of 2-4 than the flux we actually assumed
in the simulations, we make an order estimation here with this evaluation.
We summarize the evaluated fluxes using Eq. (\ref{F_ion}) in Table \ref{table:UVbackground}. 
This shows that the required ionizing photon number flux is of the order of 
$\sim 10^9~{\rm photons}~{\rm cm^{-2}}~{\rm s^{-1}}$, which roughly corresponds to $J_{21}\sim 100-1000$, where $J_{21}$ is the mean intensity at the hydrogen Lyman limit frequency in units of $10^{-21}$ erg cm$^{-2}$ s$^{-1}$ Hz$^{-1}$ sr$^{-1}$. 
This value seems to be much higher than $J_{21}$ expected for the global UV background radiation during the epoch of reionization. Thus, we consider the possibilities of local sources.  

The first possibility is Population III (Pop III) stars. 
The $\Lambda$CDM cosmology predicts that Pop III stars form in low-mass mini-halos 
with the masses of $\sim 10^{5-6}\solmass$, which collapse typically at $z\sim$10-30 \citep[e.g.,][]{Tegmark+97,Yoshida+03}. 
Although the initial mass spectrum of Pop III stars is still controversial, 
several theoretical studies have shown that Pop III stars are typically massive as $\sim100\solmass$ \citep[e.g.,][]{Nakamura&Umemura01,Susa+14,Hirano+14,Hirano+15}. 
Therefore, strong UV radiation can be expected in the vicinity of a Pop III halo. 
If we assume a Pop III star with the mass of $100-1000 \solmass$ and 
the ionizing photon emissivity of $10^{50-51}~\per{s}{1}$ \citep{Schaerer02}, 
the ionizing photon number flux is $\sim10^{8-9}~\per{cm}{2}~\per{s}{1}$ at 100~pc, 
which roughly corresponds to the virial radius of a mini-halo. 
Thus, the ionizing photon number flux to allow the formation of compact star clusters
can be easily accomplished if a Pop III star as massive as $>100\solmass$ forms at $\approx 100$pc
from a collapsing cloud. 
Also, the Kelvin-Helmholtz timescale of a Pop III star with  $\approx 100\solmass$
is $\sim 10^5$yr \citep{O'Shea&Norman07}, and 
therefore the star reaches the main sequence faster than the cloud contraction.
However, the lifetime of a Pop III star  is a few $10^6$yr. Thus, the formation of GCs by
Pop III radiation is realized only for clouds contracting within $10^6$yr.
We note that the ionizing photon number flux possibly changes with time according to the stellar motion,
if the Pop III star formation takes place during the hierarchical merging process \citep[e.g.,][]{Johnson+08}. 
The variation timescale of UV radiation is thought to be roughly the infall timescale in the GC-host halo, which is  $\sim 100~$Myr.
Since this timescale is longer than the cloud contraction time, the variation of UV radiation due to the virial motion
does not affect the cloud evolution.

The second possibility is young star-forming galaxies, e.g., Lyman $\alpha$ emitters (LAEs). 
\citet{Wise&Cen09} have numerically simulated the high-$z$ young dwarf galaxies and traced the star formation histories. 
They have shown that the starburst rises up within a few times 10 Myr and the burst-phase continues for $\sim100~$ Myr,
if the virial masses of the halo are as massive as $10^9~\solmass$. 
Although the SFR varies with the timescale of $\leq 10~$Myr,  
the luminosity changes are within a factor of three. 
Hence, the timescale condition for the formation of compact star clusters is likely to be satisfied. 
Recently, \citet{Yajima+14} have calculated the emissivities of ionizing photons of young star-forming galaxies. 
According to their result, the ionizing photon number emissivities of the galaxies at $z>6$ correspond to $\sim 10^{52-53}\: \per{s}{1}$,which is translated into the ionizing photon number flux of $\sim 10^{9} ~\per{cm}{2}~\per{s}{1}$ at 1 kpc from the galactic center. 
Thus, if the star forming regions in LAEs are as compact as $\sim 1$~kpc, 
compact star clusters may form in sub-halos of the LAEs. 

The third possibility is active galactic nuclei (AGNs). 
Recent studies have pointed out the possibility that high-$z$ quasars 
and faint AGNs bring large contribution to cosmic reionization \citep{Glikman+11,Giallongo+15,Madau&Haardt15,Yoshiura16}. 
Therefore, it is reasonable to consider UV radiation from AGNs. 
As for faint AGNs, their typical luminosity is $10^{43}$ erg/s in the range of 2-10 keV \citep{Giallongo+15}. 
If we assume a simple power-law of the spectrum energy distribution as $L_\nu \propto \nu^{-1}$, 
the ionizing photon number emitted by the AGN $\dot{N}_{\rm ion}$ is roughly estimated as $\sim 10^{53}\;\per{s} {1}$. 
Thus, even a faint AGN provides the ionizing photon number flux at 1~kpc away as  $\gtrsim 10^9 \: {\rm photons}\: \per{cm}{2}\: \per{s}{1}$. 
Several authors have argued that the duty cycle of the AGN activity is
in the timescale of $10^{8}~$Myr \citep[e.g.,][]{Haehneit+98},
which is comparable to the Eddington timescale. 
If the mass accretion on to a central black hole is driven by a nuclear starburst,
the accretion timescale can be as short as $10^7$yr \citep{Umemura+97}.
If this gives the rise time of luminosity, then the situation is favorable for the ``supersonic infall''. 
Therefore,  we can expect the formation of  compact star clusters, 
if the rise time of the AGN luminosity is shorter than 10~Myr.

\subsection{Effect of Tidal Field}
As shown in \S~\ref{sec:comparison},  
low-mass ($\lesssim 10^5~\solmass$) GCs are not formed in our simulations.
Here, we assess the effect of tidal stripping by host galaxies, which might work so as to reduce the masses of GCs. 
Assuming a host galaxy as a point-mass for simplicity, 
the tidal radius $r_{\rm t}$  of a star cluster orbiting a host galaxy is roughly given by 
\begin{equation}
	\label{eq:rt_difinition}
	\frac{Gm(r<r_{\rm t})}{r_{\rm t}^2} \sim 2\frac{GM_{\rm gal}m(r<r_{\rm t})r_{\rm t}}{r_{\rm gal}^3}, 
\end{equation}
where $m(r<r_{\rm t})$, $M_{\rm gal}$, and $r_{\rm gal}$ denote the cumulative cluster mass within the tidal radius $r_{\rm t}$, the host galaxy mass, and the distance from the galactic center to the cluster, respectively. 
Supposing a high-$z$ low-mass galaxy of $M_{\rm gal} = 10^9\solmass$ and $r_{\rm gal}=$ 0.3-1 kpc and using the simulated mass profiles (Fig.~\ref{fig:cumulative_mass}), the tidal radii are estimated to be a few $\times$ 10 pc to $\sim$ 100 pc. 
This estimation implies that the star-dominant parts of the compact star clusters likely to gradually lose their masses in the tidal fields according as the two-body relaxation proceeds, while the diffuse dark matter components would be totally stripped away as shown by \citet{Saitoh+06}. 
Furthermore, the variety of the orbits of star clusters possibly leads to the variety of mass-loss rates of the clusters. 
Hence, it seems important to take the tidal stripping into consideration 
for more quantitative comparison between simulations and observations.  

\subsection{Internal feedback processes}
Throughout this paper, we have concentrated on the impacts of the external background radiation
but neglected internal feedback processes. 
Actually, stars formed in self-shielding regions are expected to emit UV radiation, which ionizes the self-shielded regions internally. 
Besides, type II supernova (SN) explosions pose dynamical impacts on the gas in the star forming regions. 
These internal feedbacks may play a significant role to regulate the subsequent star formation \citep[e.g.,][]{Kitayama+04,Kitayama&Yoshida05,Hasegawa&Semelin13}. 
  
\citet{Kitayama+04} have explored the impact of the internal UV radiation feedback by a massive Pop III star in a low-mass halo with $10^6\solmass$, and found that the feedback reduces the ambient gas density by photo-evaporation and suppresses the subsequent star formation. 
In our simulations, the mass resolution is $\sim 10^3\solmass$ (\S~\ref{sec:method}), which correspond to
the mass of stars formed simultaneously. 
Then, the emitted ionizing photon number is evaluated as $\sim 10^{50}~\per{s}{1}$ 
by utilizing STARBURST99 \citep{STARBURST99} assuming an instantaneous starburst model 
for $Z/Z_{\odot} = 0.02$ and the Salpeter IMF. 
Therefore, the argument by \cite{Kitayama+04} is partially applicable to our simulations, 
and the subsequent star formation in the gas clouds is expected to be suppressed by the internal UV feedback. 
However, in the present situation, UV radiation from stars surrounding the cloud center 
might positively work to compress the central star-forming region. 
Since such a complicated behavior is expected, it is hard to assess how much the internal UV feedback 
quantitatively affects our results, before the internal UV feedback is actually incorporated. 
On the other hand, we expect that SN feedbacks are unlikely to affect on the formation of compact star clusters, 
since the star formation is quickly quenched by $\sim 5$Myr (see Fig.~\ref{fig:SFR}).

\section{Summary}
In this paper, we have performed three-dimensional radiation hydrodynamic simulations 
to explore the star cluster formation under UV background radiation. 
In particular, we have paid attention on three-dimensional effects that were not studied in the previous 1D-RHD calculations by HUK09. 
We have shown that low-mass clouds with masses of $2.5\times10^6-10^7\solmass$ 
can collapse to form stars even under strong UV background radiation if they are contracting with supersonic velocities.
As a result, we have demonstrated that the mechanism proposed by HUK09 does work, even if three-dimensional effects 
are incorporated.   

In the case that UV background radiation is extremely anisotropic, i.e., a cloud is irradiated 
by UV radiation from one side, shaded neutral regions emerge on the opposite side of the UV source. 
However, the shaded regions are shoved by surrounding UV heated gas towards the center of the cloud.
Consequently, the gas components are converted to the stellar component at the central compact regions 
in the cloud ($\sim 10$ pc) ,  regardless of the anisotropy of the background UV radiation.  
Although the anisotropy of the background radiation slightly affects the star formation histories, 
the duration of star formation becomes as short as $\sim 10$ Myr and the stellar age dispersion is less than 10 Myr. 
Hence, the star clusters formed via ``supersonic infall'' tend to be of the single stellar population, 
which is favorable to reproduce the feature of GCs. 

We have pursued the stellar dynamics of simulated star clusters, and shown 
that the star clusters formed via ``supersonic infall'' become compact, stellar-dominated systems,
owing to the compactness of self-shielded regions as well as the strong dissipation of the cloud contraction energy. 
We have found that the half-mass radius, mass-to-light ratio, and velocity dispersion of simulated star clusters 
are similar to observed GCs. 
The results are not affected by the time evolution of the background UV intensity, 
as long as the rise time of UV intensity is shorter than 10~Myr. 
We have also confirmed that no GC-like star clusters can form if the ``prompt star formation'' occurs. 
In the ``prompt star formation'', stars begin to form at an earlier phase of the cloud contraction 
due to the prompt self-shielding and the formation of abundant $\HH$ molecules around an ionization front. 
As a consequence of the weak energy dissipation, the star clusters become diffuse, dark matter-dominated systems,
which are clearly distinguished from GCs. 
Hence, we conclude that the intensity of UV background radiation significantly regulates 
the properties of the star clusters, and the ``supersonic infall'' under a strong UV background seems to be 
a potential scenario for the formation of GCs, as proposed in HUK09.

\section*{Acknowledgements}
We are grateful to M. Mori, K. Yoshikawa, A. Y. Wagner, A. Burkert, F. Nakamura, H. Yajima and A. Inoue for fruitful discussion. 
The numerical simulations have been performed with COMA provided by Interdisciplinary 
Computational Science Program in Center for Computational Sciences, University of Tsukuba, 
and with Cray XC30 at Center for Computational Astrophysics, NAOJ. 
This research was supported in part by Grant-in-Aid for Scientific Research (B) 
by JSPS (15H03638) (MU) and a grant from NAOJ (KH).




\bibliographystyle{mnras}
\bibliography{mn_main}



%
%


\bsp	
\label{lastpage}
\end{document}